\documentclass[12pt,a4paper]{article}

\usepackage{amsmath}
\usepackage{amsfonts}
\usepackage{amssymb}
\usepackage{graphicx}
\usepackage{color}
\usepackage{booktabs}
\usepackage{inputenc}
\usepackage[T1]{fontenc}
\usepackage{mathrsfs}
\usepackage{enumerate}
\usepackage{xcolor,url}
\usepackage{cancel,dsfont} 

\def\hinvMpc{h\,{\rm Mpc}^{-1}}

\newcommand{\half}{{\textstyle{\frac12}}}

\newcommand{\kmax}{k_{\rm max }}

\newcommand{\code}[1]{\texttt{#1}}

\usepackage[colorlinks,bookmarks]{hyperref}
\definecolor{linkblue}{rgb}{0,0,0.8}
\definecolor{linkgreen}{rgb}{0,0.5,0}

\hypersetup{pdfpagemode=UseNone, pdfstartview=FitH, linkcolor=linkblue,
 citecolor=linkgreen, urlcolor=linkblue}

\newcommand{\bea}{\begin{eqnarray}}
\newcommand{\eea}{\end{eqnarray}}
\newcommand{\be}{\begin{equation}}
\newcommand{\ee}{\end{equation}}
\newcommand{\fr}[2]{\frac{ #1}{#2}}

\newcommand{\cH}{\mathcal{H}}

\newcommand{\knl}{k_{\rm NL}}

\newcommand{\barr}{\begin{array}}
	\newcommand{\earr}{\end{array}}

\newcommand{\sfrac}[2]{{\textstyle\frac{#1}{#2}}}

\newcommand{\mpl}{M_{\rm Pl}}

\newcommand{\css}{c_s^2}

\newcommand{\xvec}{\vec{x}}

\newcommand{\rhod}{ \rho_D}
\newcommand{\rhom}{ \rho_m} 
\newcommand{\rhomi}{ \rho_{m,0}} 
\newcommand{\rhodi}{ \rho_{D,0}}

\newcommand{\deltagu}{\delta g^{00}_{\rm u}}

\newcommand{\ta}{\tilde{a}}

\newcommand{\mG}{\mathcal{G}}

\newcommand{\mU}{\mathcal{U}}
\newcommand{\mV}{\mathcal{V}}

\newcommand{\vk}{\vec{k}}
\newcommand{\vq}{\vec{q}}

\newcommand{\vx}{\vec{x}}

\def\beq{\begin{equation}}
\def\eeq{\end{equation}}
\def\be{\begin{equation}}
\def\ee{\end{equation}}
\def\bea{\begin{eqnarray}}
\def\eea{\end{eqnarray}}
\def\d{{\partial}}

\def\mpl{M_{\rm Pl}}
\def\nn{\nonumber}

\def\k{{\bf k}}
\def\knl{{k_{\rm NL}}}
\def\km{{k_{\rm M}}}

\def\xfl{{\vec x_{\rm fl}}}

\newcommand{\q}{\vec{q}}

\newcommand{\Comment}[1]{{}}

\begin{document}

\begin{center}

{\Large \bf {Limits on Clustering {and Smooth} Quintessence\\[0.3cm]from the EFTofLSS}
}
\\[0.7cm]

{\large Guido D'Amico${}^{1,2}$, Yaniv Donath${}^{3,4,5}$, Leonardo Senatore${}^{4,5}$, Pierre Zhang${}^{6,7,8}$
\\[0.7cm]}
\end{center}

\begin{center}

\vspace{.0cm}

{\normalsize { \sl $^{1}$ University of Parma, Department of Mathematical, Physical and Computer Sciences,\\ Parma, Italy}}\\
\vspace{.3cm}

{\normalsize { \sl $^{2}$ INFN Gruppo Collegato di Parma, Parma, Italy}}\\
\vspace{.3cm}

{\normalsize { \sl $^{3}$ Institute for Theoretical Physics, ETH Zurich, 8093 Zurich, Switzerland}}\\
\vspace{.3cm}

{\normalsize { \sl $^{4}$ Stanford Institute for Theoretical Physics, Physics Department,\\ Stanford University, Stanford, CA 94306}}\\
\vspace{.3cm}

{\normalsize { \sl $^{5}$ 
Kavli Institute for Particle Astrophysics and Cosmology,\\
 SLAC and Stanford University, Menlo Park, CA 94025}}\\
\vspace{.3cm}

{\normalsize { \sl $^{6}$ Department of Astronomy, School of Physical Sciences, \\
University of Science and Technology of China, Hefei, Anhui 230026, China}}\\
\vspace{.3cm}

{\normalsize { \sl $^{7}$ CAS Key Laboratory for Research in Galaxies and Cosmology, \\
University of Science and Technology of China, Hefei, Anhui 230026, China}}\\
\vspace{.3cm}

{\normalsize { \sl $^{8}$ School of Astronomy and Space Science, \\
University of Science and Technology of China, Hefei, Anhui 230026, China}}\\
\vspace{.3cm}


\end{center}

\hrule \vspace{0.3cm}
{\small \noindent \textbf{Abstract} We apply the Effective Field Theory of Large-Scale Structure (EFTofLSS) to analyze cosmological models with clustering quintessence, {which allows us to consistently describe the parameter region in which the quintessence equation of state $w < - 1$}.
First, we extend the description of biased tracers in redshift space to the presence of clustering quintessence, and compute the one-loop power spectrum. We solve the EFTofLSS equations using the exact time dependence, which is relevant to obtain unbiased constraints. Then, fitting the full shape of BOSS pre-reconstructed power spectrum measurements, the BOSS post-reconstruction BAO measurements, BAO measurements from 6DF/MGS and eBOSS, the Supernovae from Pantheon, and a prior from BBN, we bound the {clustering} quintessence equation of state parameter $w=-1.011_{-0.048}^{+0.053}$ at $68\%$ C.L..
Further combining with Planck, we obtain $w=-1.028_{-0.030}^{+0.037}$ at $68\%$ C.L..
{We also obtain constraints on {smooth} {quintessence, in the physical regime $w \geq -1$: combining all datasets, we get $-1\leq w < - 0.979$ at $68\%$ C.L..} {These results strongly support a cosmological constant}.
\vspace{0.3cm}}
\hrule

\newpage

\tableofcontents

\section{Introduction and Summary\label{sec:intro}}

\paragraph{Introduction} 
The analysis of the Full Shape (FS) of the BOSS galaxy power spectrum with the Effective Field Theory of Large-scale Structure (EFTofLSS) at one loop has provided us with a measurement of all parameters in $\Lambda$CDM with just a Big Bang Nucleosynthesis~(BBN) prior~\cite{DAmico:2019fhj,Ivanov:2019pdj,Colas:2019ret} (see also~\cite{Philcox:2020xbv} for other prior choices and~\cite{DAmico:2019fhj} for a joint analysis with the BOSS bispectrum using the tree-level prediction). 
The combination with BOSS reconstructed measurements and baryon acoustic oscillations (BAO) from eBOSS, as well as with supernovae redshift-distance or cosmic microwave background~(CMB) measurements, has further allowed us to bound the total neutrino mass, and put limits on the effective number of relativistic species, on smooth dark energy, or on curvature~\cite{DAmico:2019fhj,Colas:2019ret,Ivanov:2019hqk,Philcox:2020vvt,DAmico:2020kxu,Chudaykin:2020ghx}. 
In particular, the FS analysis can help constrain models invented to address the Hubble tension as it provides measurements independent on the CMB or local distance ladders~\cite{DAmico:2020ods,Ivanov:2020ril,Niedermann:2020qbw,Smith:2020rxx}.

All these results were made possible thanks to the development of the EFTofLSS, which is a powerful tool to extract cosmological information from Large-Scale Structure surveys. 
A long line of study was necessary to bring the framework to the level where it can be applied to the data. 
We therefore find fair to add the following footnote where we acknowledge a fraction of its important developments, though not all intermediate results are used in the present analysis~\footnote{The initial formulation of the EFTofLSS was performed in Eulerian space in~\cite{Baumann:2010tm,Carrasco:2012cv}, and subsequently extended to Lagrangian space in~\cite{Porto:2013qua}. 
The dark matter power spectrum has been computed at one-, two- and three-loop orders in~\cite{Carrasco:2012cv, Carrasco:2013sva, Carrasco:2013mua, Carroll:2013oxa, Senatore:2014via, Baldauf:2015zga, Foreman:2015lca, Baldauf:2015aha, Cataneo:2016suz, Lewandowski:2017kes,Konstandin:2019bay}.
These calculations were accompanied by some theoretical developments of the EFTofLSS, such as a careful understanding of renormalization~\cite{Carrasco:2012cv,Pajer:2013jj,Abolhasani:2015mra} (including rather-subtle aspects such as lattice-running~\cite{Carrasco:2012cv} and a better understanding of the velocity field~\cite{Carrasco:2013sva,Mercolli:2013bsa}), of several ways for extracting the value of the counterterms from simulations~\cite{Carrasco:2012cv,McQuinn:2015tva}, and of the non-locality in time of the EFTofLSS~\cite{Carrasco:2013sva, Carroll:2013oxa,Senatore:2014eva}.
These theoretical explorations also include enlightening studies in 1+1 dimensions~\cite{McQuinn:2015tva,Pajer:2017ulp}.
An IR-resummation of the long displacement fields had to be performed in order to reproduce the Baryon Acoustic Oscillation (BAO) peak, giving rise to the so-called IR-Resummed EFTofLSS~\cite{Senatore:2014vja,Baldauf:2015xfa,Senatore:2017pbn,Lewandowski:2018ywf,Blas:2016sfa}.
An account of baryonic effects was presented in~\cite{Lewandowski:2014rca,Braganca:2020nhv}. The dark-matter bispectrum has been computed at one-loop in~\cite{Angulo:2014tfa, Baldauf:2014qfa}, the one-loop trispectrum in~\cite{Bertolini:2016bmt}, and the displacement field in~\cite{Baldauf:2015tla}.
The lensing power spectrum has been computed at two loops in~\cite{Foreman:2015uva}.
Biased tracers, such as halos and galaxies, have been studied in the context of the EFTofLSS in~\cite{ Senatore:2014eva, Mirbabayi:2014zca, Angulo:2015eqa, Fujita:2016dne, Perko:2016puo, Nadler:2017qto} (see also~\cite{McDonald:2009dh}), the halo and matter power spectra and bispectra (including all cross correlations) in~\cite{Senatore:2014eva, Angulo:2015eqa}. Redshift space distortions have been developed in~\cite{Senatore:2014vja, Lewandowski:2015ziq,Perko:2016puo}.
Neutrinos have been included in the EFTofLSS in~\cite{Senatore:2017hyk,deBelsunce:2018xtd}, clustering dark energy in~\cite{Lewandowski:2016yce,Lewandowski:2017kes,Cusin:2017wjg,Bose:2018orj}, and primordial non-Gaussianities in~\cite{Angulo:2015eqa, Assassi:2015jqa, Assassi:2015fma, Bertolini:2015fya, Lewandowski:2015ziq, Bertolini:2016hxg}. 
The exact-time dependence in the loop has been clarified in~\cite{Donath:2020abv,Fujita:2020xtd}. 
Faster evaluation schemes for the calculation of some of the loop integrals have been developed in~\cite{Simonovic:2017mhp}.
Comparison with high-fidelity $N$-body simulations to show that the EFTofLSS can accurately recover the cosmological parameters have been performed in~\cite{DAmico:2019fhj,Colas:2019ret,Nishimichi:2020tvu}.}. 

In this paper, we analyze the BOSS FS power spectrum using the EFTofLSS at one loop in the context of clustering quintessence~\cite{Creminelli:2006xe,Creminelli:2008wc,Creminelli:2009mu} {and smooth quintessence}. 
In clustering quintessence, dark energy is made of a scalar field (the quintessence field) whose fluctuations have {effectively} zero speed of sound, $c_s$, and therefore `cluster', as they can fall into gravitational potentials. 
It is a particularly appealing model since the dark energy equation of state parameter $w$ can cross the so-called phantom divide, {$w=-1$ and consistently describe the regime $w<-1$. }
This is allowed thanks to the presence of higher-derivative operators in the Lagrangian that stabilize gradient instabilities, but this can only happen if $c_s^2 \rightarrow 0$ such that they remain not parametrically suppressed. 
Clustering quintessence has been considered within the context of structure formation in~\cite{Sefusatti:2011cm,DAmico:2011pf,Anselmi:2011ef,Anselmi:2014nya} and in the EFTofLSS in~\cite{Lewandowski:2016yce} (see also~\cite{Lewandowski:2017kes,Cusin:2017wjg,Bose:2018orj} for embeddings of other dark energy theories in the EFTofLSS). 
In this work, we extend the description to biased tracers in redshift space with exact-time dependence in order to apply it to data from galaxy surveys. {We remark that we find it quantitatively important to solve the EFTofLSS equations with the exact time dependence, rather than with the approximate, so-called `EdS', approximation.} {As for smooth quintessence, which has already been analyzed in light of the BOSS FS and LSS data in~\cite{DAmico:2020kxu}, we will perform here the analysis by imposing a physical flat prior $-1 \leq w$ on the smooth quintessence equation of state parameter. 
By $w$CDM, we refer to a Universe that includes a smooth dark energy component, i.e. a scalar quintessence field with $c_s^2 \rightarrow 1$, whose perturbations can be neglected since the sound horizon is of the size of the cosmological horizon.
In this picture, $w<-1$ is an unphysical region where the vacuum is unstable, therefore we should analyze $w$CDM excluding this region (see discussions in e.g.~\cite{Cline:2003gs,Creminelli:2006xe}). }

This paper is organized as follows. 
We compute the power spectrum at one loop in redshift space for biased tracers with exact time dependence for the clustering quintessence model in Section~\ref{sec:theory}.
Further details concerning this derivation are given in the appendices. 
In Section~\ref{sec:analysis}, we apply our framework to LSS data, and in appendix~\ref{appendixd} we show the full posteriors including nuisance parameters.
Our results are summarized at the end of this Introduction.

\paragraph{Data sets} We analyze the FS of BOSS DR12 pre-reconstructed power spectrum measurements~\cite{Gil-Marin:2015sqa}, baryon acoustic oscillations (BAO) of BOSS DR12 post-reconstructed power spectrum measurements~\cite{Gil-Marin:2015nqa}, 6DF~\cite{Beutler:2011hx} and SDSS DR7 MGS~\cite{Ross:2014qpa}, as well as high redshift Lyman-$\alpha$ forest auto-correlation and cross-correlation with quasars from eBOSS DR14 measurements~\cite{Agathe:2019vsu, Blomqvist:2019rah}.
We also consider combinations with Supernovae (SN) measurements from the Pantheon sample~\cite{Scolnic:2017caz} and with Planck2018 TT,TE,EE+lowE+lensing~\cite{Aghanim:2018eyx}. 

\paragraph{Methodology} We analyze the BOSS FS using the galaxy power spectrum in redshift space at one loop in the EFTofLSS~\cite{Perko:2016puo} following the methodology described in~\cite{DAmico:2019fhj,Colas:2019ret}. 
The description of the likelihood, including the covariances and priors used, can be found in~\cite{DAmico:2019fhj}.
The theory of biased tracers in redshift space with exact time dependence in clustering quintessence cosmology at one loop is derived in Section~\ref{sec:theory} {(see also~\cite{Fujita:2020xtd} which has already derived the same expressions, but just in real space, with a different approach)}, and the scale cut up to which the FS is analyzed is discussed in Sec.~\ref{sec:scalecut}.
The power spectrum is IR-resummed~\cite{Senatore:2014vja,Senatore:2017pbn,Lewandowski:2018ywf,DAmico:2020kxu}, and includes corrections to observational systematics: the Alcock-Paczynski effect~\cite{Alcock:1979mp}, window functions~\cite{Beutler:2018vpe}, and fiber collisions~\cite{Hahn:2016kiy}. 

We sample over the following cosmological parameters: the abundance of baryons $\omega_b$, the abundance of cold dark matter $\omega_{\rm cdm}$, the Hubble constant $H_0$, the amplitude of the primordial fluctuations $\ln (10^{10} A_s)$, the tilt of the primordial power spectrum $n_s$, and the quintessence equation of state parameter $w$.
We impose no prior on the cosmological parameters but a BBN prior on $\omega_b$: a Gaussian prior centered on $0.02235$ with $\sigma_{\rm BBN} = 0.0005$, obtained by adding up the theory and statistical errors of~\cite{Cooke:2017cwo}. 
We use the Planck prescription of one single massive neutrino with mass $0.06$ eV as done in~\cite{Aghanim:2018eyx}. 
Allowing the EFT parameters to vary only within physical ranges, we impose priors on them as in~\cite{DAmico:2020kxu}. 

The BAO measurements from the post-reconstructed BOSS power spectrum are correlated with BOSS pre-reconstructed (FS) measurements. 
The joint analysis is described in~\cite{DAmico:2020kxu} (see also~\cite{Philcox:2020vvt}). 
When adding BAO from 6DF/MGS or eBOSS, SN from Pantheon, or Planck data, we simply add the log-likelihoods as these measurements are uncorrelated among each other.
We neglect the small cross-correlation between LSS data with Planck weak lensing and the integrated Sachs-Wolfe (ISW) effect. 

\begin{figure}
\centering
\includegraphics[width=0.76\textwidth]{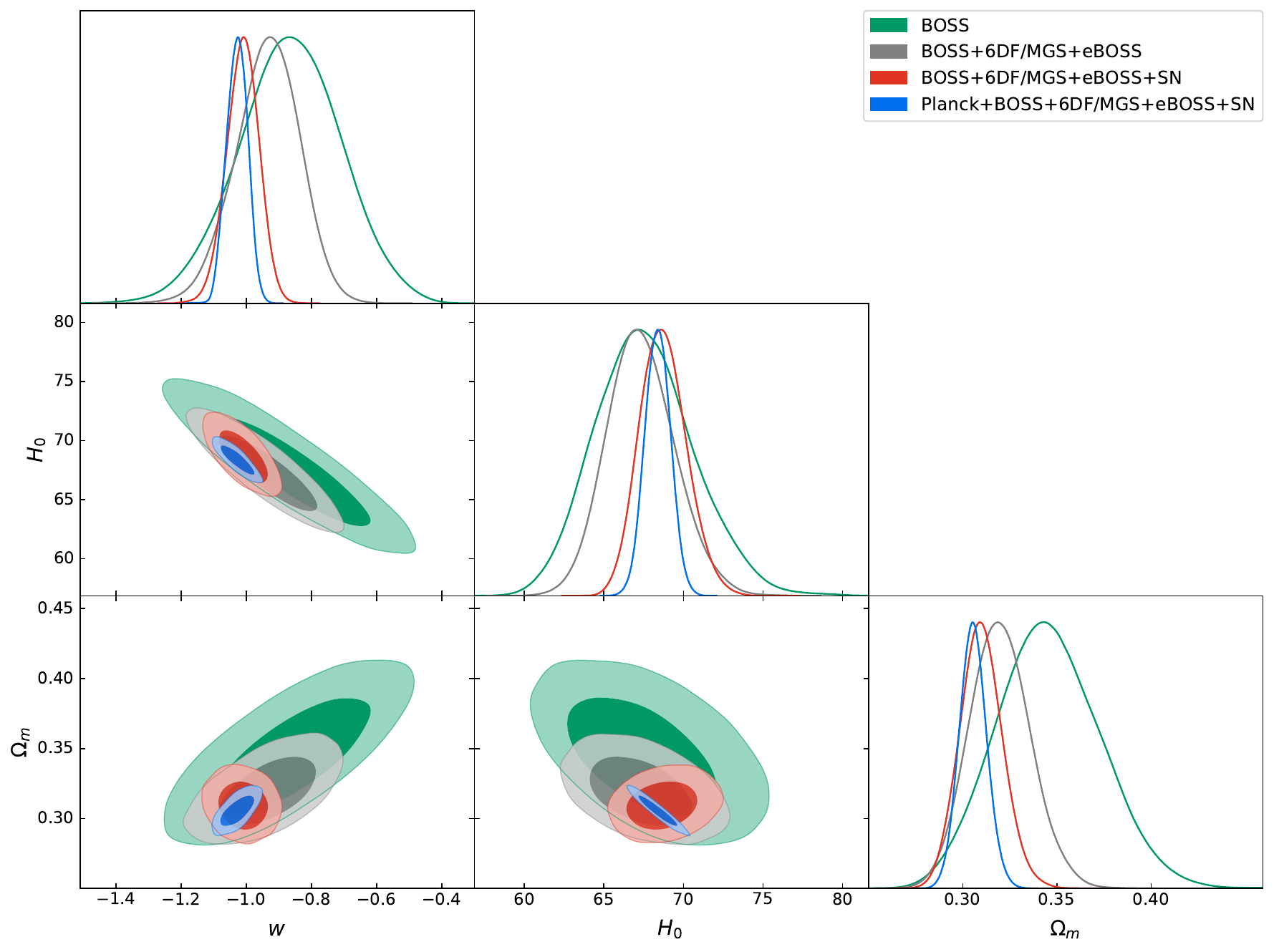}
\caption{\small 1D and 2D posteriors of $w$, $H_0$ and $\Omega_m$ in clustering quintessence from various analyses performed in this work. When not analyzed in combination with Planck, we use a BBN prior. 
} 
\label{fig:summary_probes}
\end{figure}

\paragraph{Main Results} Our main results are best represented by Fig.~\ref{fig:summary_probes}. 
Fitting BOSS FS + BOSS reconstructed BAO with a BBN prior on clustering quintessence, we are able to measure all cosmological parameters. 
In particular, we determine the quintessence equation of state parameter $w$, the present matter fraction $\Omega_m$, and the Hubble constant $H_0$, to $18\%, 8.2\%$, and $4.6\%$ precision, respectively, at $68\%$ confidence level (C.L.), finding $w=-0.867_{-0.15}^{+0.17}$, $\Omega_m=0.3456_{-0.027}^{+0.03}$, and $H_0=67.58_{-3.5}^{+2.7}$. {We also determine $\ln (10^{10}A_{s })=2.64_{-0.17}^{+0.16}$ and $n_{s }=0.8884_{-0.059}^{+0.072}$ at 68\% C.L..} 
Upon addition of the BAO measurements from 6DF/MGS and eBOSS, and SN measurements from Pantheon, we get $w=-1.011_{-0.048}^{+0.053}$, $\Omega_m=0.3099_{-0.011}^{+0.012}$, and $H_0=68.72_{-1.6}^{+1.4}$, which amounts to error bar reductions of $68\%, 60\%$, and $52\%$, respectively. {We also find $\ln (10^{10}A_{s })=2.806_{-0.16}^{+0.15}$ and $n_{s }=0.9335_{-0.05}^{+0.054}$ at 68\% C.L.}.
Adding Planck data (Table~\ref{tab:qcmb}), we finally constrain $w$, $\Omega_m$, and $H_0$ to $3.3\%, 2.4\%$, and $1.2\%$ precision, respectively, obtaining $w=-1.028_{-0.030}^{+0.037}$, $\Omega_m=0.3055_{-0.0073}^{+0.0074}$, and $H_0=68.38_{-0.84}^{+0.78}$ {, and also obtain $\ln (10^{10}A_{s })=3.046_{-0.014}^{+0.014}$ and $n_{s }=0.9665_{-0.0036}^{+0.0042}$ at 68\% C.L..} 

All analyses performed here show that our Universe is consistent with $\Lambda$CDM.
First, clustering quintessence in the limit $w=-1$ reduces to $\Lambda$CDM, and we find that $w$ is consistent with $-1$ at $\lesssim 68\%$ C.L.
Second, the values obtained for the other cosmological parameters in clustering quintessence are consistent within $68\%$ C.L. with the $\Lambda$CDM ones obtained by fitting BOSS FS with the EFTofLSS~\cite{DAmico:2019fhj,Ivanov:2019pdj,Colas:2019ret}, in combination with other probes~\cite{Ivanov:2019hqk,Philcox:2020vvt,DAmico:2020kxu}, or fitting Planck alone~\cite{Aghanim:2018eyx}~\footnote{With an exception on $\ln (10^{10} A_s)$ which is consistent at $\sim 2\sigma$ with Planck.}. 

A similar observation applies when fitting $w$CDM with a flat prior on the dark energy equation of state parameter of $w\geq-1$ (Table~\ref{tab:wcdm}).
Fitting BOSS data with a BBN prior, we find in this case $\Omega_m=0.337_{-0.022}^{+0.017}$ and $H_0=68.6\pm 1.8$, and we bound $-1\leq w < -0.91$ {at $68\%$ C.L. ($-1 \leq w < -0.81$ at $95\%$ C.L.)}. {We also get $\ln (10^{10}A_{s })=2.77\pm0.19$ and $n_{s }=0.885_{-0.058}^{+0.069}$ at 68\% C.L..}
Adding BAO measurements, Pantheon SN and Planck data we obtain the very stringent constraint $-1\leq w< -0.979$ {at $68\%$ C.L. ($-1 \leq w < -0.956$ at $95\%$ C.L.)}.
Thus, allowing $w$CDM only within the physical region gives {tight} posteriors that are also consistent with the ones obtained on $\Lambda$CDM fitting BOSS or Planck. 
This is illustrated in Fig.~\ref{fig:summary_qwcdm}. 

We end this summary of the main results with a note of warning. It should be emphasized that in performing this analysis, as well as the preceding ones using the EFTofLSS by our group~\cite{DAmico:2019fhj,Colas:2019ret,DAmico:2020kxu,DAmico:2020ods}, we have assumed that the observational data are not affected by any unknown systematic error, such as, for example, line of sight selection effects or undetected foregrounds. In other words, we have simply analyzed the publicly available data for what they were declared to be: the power spectrum of the galaxy density in redshift space. Given the additional cosmological information that the theoretical modeling of the EFTofLSS allows us to exploit in BOSS data, it might be worthwhile to investigate if potential undetected systematic errors might affect our results. We leave an investigation of these issues to future work.

\begin{figure}[h]
\centering
\includegraphics[width=0.76\textwidth]{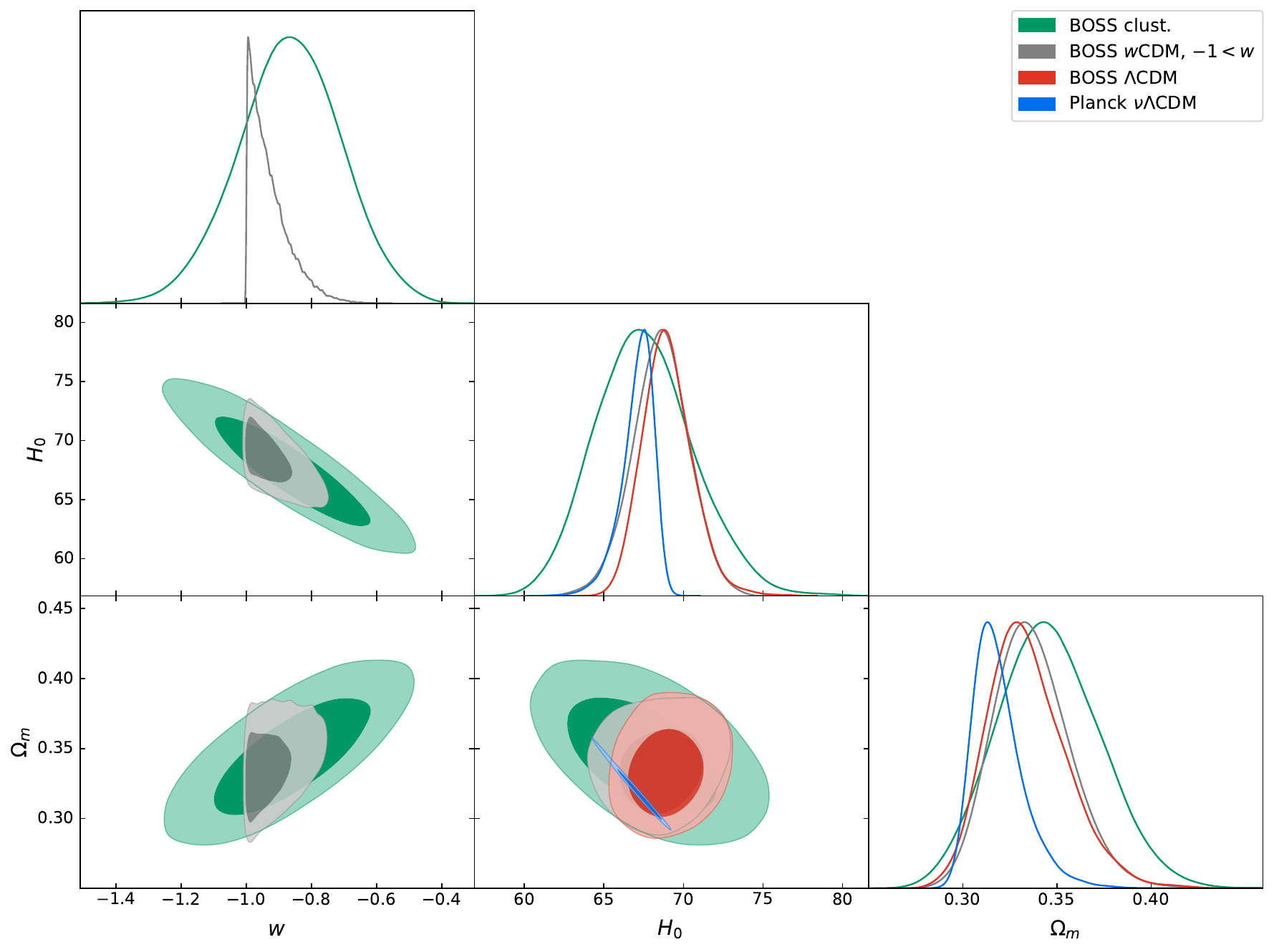}
\caption{\small 1D and 2D posteriors of $w$, $H_0$ and $\Omega_m$ obtained by fitting clustering quintessence, $w$CDM or $\Lambda$CDM to BOSS with a BBN prior. 
For $w$CDM, i.e. smooth quintessence, we restrict to the physical region $w\geq -1$. 
For comparison, we show the contours of Planck obtained in $\Lambda$CDM in the presence of massive neutrinos. The neutrinos introduce additional degeneracies in the $\Omega_m-H_0$ plane in the CMB analysis. On the contrary, fixing the neutrinos when analyzing BOSS does not significantly change the constraints {on the shown cosmological parameters}, see e.g. Table 2 in~\cite{Colas:2019ret}. This plot illustrates the consistency of the datasets as well as the consistency of the present analyses with a cosmological constant. 
} 
\label{fig:summary_qwcdm}
\end{figure}

\paragraph{Public Code} The redshift-space one-loop galaxy power spectra in the EFTofLSS are evaluated using PyBird: Python code for Biased tracers in ReDshift space~\cite{DAmico:2020kxu}~\footnote{\href{https://github.com/pierrexyz/pybird}{https://github.com/pierrexyz/pybird}}.
The exact time dependence and the clustering quintessence modifications are publicly available in PyBird.
The linear power spectra are evaluated with the CLASS Boltzmann code~\cite{Blas_2011}~\footnote{\href{http://class-code.net}{http://class-code.net}}.
The posteriors are sampled using the MontePython cosmological parameter inference code~\cite{Brinckmann:2018cvx, Audren:2012wb}~\footnote{\href{https://github.com/brinckmann/montepython\_public}{https://github.com/brinckmann/montepython\_public}}.
The triangle plots are obtained using the GetDist package~\cite{Lewis:2019xzd}.

\section{Biased tracers with exact time dependence in clustering quintessence}\label{sec:theory}

In this section, we extend the study of biased tracers in redshift space with exact time dependence, first studied in \cite{Donath:2020abv,Fujita:2020xtd}, to clustering quintessence.

\subsection{Review of the EFTofLSS with clustering quintessence}
We start by reviewing the underlying equations of motion for dark matter and the dark energy component.
For a more detailed discussion, we refer the reader to~\cite{Lewandowski:2016yce}.
In the EFT of dark energy, previously studied in \cite{Creminelli:2006xe,Cheung:2007st,Creminelli:2008wc,Gubitosi:2012hu}, the dark energy degree of freedom is assumed to be the Goldstone boson arising from the spontaneous breaking of time diffeomorphisms.
To write the most general theory, we work in unitary gauge where the scalar degree of freedom appears in the metric.
The gravitational action will contain operators that break time diffeomorphisms, while remaining invariant under time-dependent spatial diffeomorphisms.
Up to second order in perturbations, and at leading order in the derivatives, the action reads
\be \label{fullaction}
S_G = \int d^4 x \, \sqrt{-g}\bigg[ \frac{\mpl^2}{2} R - \Lambda(t) - c(t) g_{\rm u}^{00} + \frac{M_2^4(t)}{2} ( \delta g_{\rm u}^{00} )^2 
- \frac{\bar{m}_1^3}{2} \delta g_{\rm u}^{00} \delta K_{\rm u}
- \frac{\hat{m}_1^2}{2} \delta K_{\rm u}^2
- \frac{\hat{m}_2^2}{2} \delta K_{ {\rm u}, ij} \delta K^{ij}_{\rm u} \bigg] \, ,
\ee
where we use the `u' subscript, to emphasize that the metric in the action above is in unitary gauge.
Here $g_{\rm u}^{00} = -1 + \delta g_{\rm u}^{00}$ is the $00$ component of the (inverse) metric, and $\Lambda(t)$, $c(t)$, $M_2(t)$, $\bar{m}_1(t)$, $\hat{m}_1(t)$, $\hat{m}_2(t)$ are coefficients which depend on the background evolution.
Then $\delta K_{ij}$ is the perturbation of the extrinsic curvature tensor, and $\delta K$ is its trace.
For simplicity, in the following we work with $\bar{m}_1 = 0$, but it can be checked~\cite{Creminelli:2006xe,Creminelli:2008wc} that this operator describes a clustering quintessence at cosmological scales.
The operators proportional to $\hat{m}_i$ are negligible on large scales as they scale as $\sim k^4$, but are necessary to guarantee the stability of perturbations, as discussed below.
To $S_G$, we add the action for matter $S_M$, which we take to be fully diffeomorphism invariant.
This guarantees that, once we explicitly reintroduce the Goldstone mode $\pi$, there will be no direct couplings of $\pi$ to matter.

The background equations we obtain from $S_G + S_M$ are the familiar Friedmann equations: 
\begin{align}
	3 H^2 \mpl^2 & = \rho_m + \rho_D \, , \\
	- 2 \dot H \mpl^2 & = \rhom + \rho_D + p_D \, ,
 \label{anotherFriedmann} 
\end{align}
where we set the cold dark matter pressure $p_m = 0$, and define the background dark energy density, $\rho_D$, and pressure, $p_D$, by
\begin{align} \label{cequation1}
c(t) & = \half \left( {\rho}_D + {p}_D \right), \\
\Lambda(t)& = \half \left( {\rho}_D - {p}_D \right) \label{lambdaequation1}\ .
\end{align} 
From the Friedmann equations we obtain the background solutions for the dark matter and dark energy densities: 
\be\label{denssol}
\rhom = \rhomi a^{-3}, \qquad \rhod = \rhodi a^{-3 (1+w)} \ ,
\ee
where the sub index ${}_0$ stands for the present day value, and we use the equation of state parameter for dark energy $w = {p}_D/ {\rho}_D $.
In the following, we will often use the present day fractional densities $\Omega_{x,0} = \sfrac{\rho_{x,0}}{\rho_{D,0}+\rho_{m,0}}$, with $x \in \{m,D\}$.

Starting from the action in unitary gauge, it is useful to explicitly reintroduce the Goldstone mode doing the Stueckelberg trick.
We perform the time diffeomorphism $ x^0 \rightarrow x^0 + \xi^0(\xvec , t)$ and $x^i \rightarrow x^i$, and then substitute $\xi^0(x) \rightarrow - \pi(x)$.
The replacement rules for the coefficients and the metric are (for details see {for example} \cite{Lewandowski:2016yce})
\begin{align}\label{metric-expand}
c(t)& \rightarrow c(t + \pi) = c(t) + \dot c(t) \pi + \half \ddot c(t) \pi^2 + \dots \, ,\\
g^{00}_{\rm u}& \rightarrow g^{00} + 2 g^{0 \mu } \partial_\mu \pi + g^{\mu \nu} \partial_\mu \pi \partial_\nu \pi \, .
\end{align}
Gravitational perturbations will be described by the spatially flat perturbed FLRW metric in Newtonian gauge:
\be\label{metric-pert}%
ds^2=-(1+2\Phi)dt^2+a(t)^2(1-2\Psi)\delta_{ij}dx^idx^j\, ,
\ee
where $\Phi$ and $\Psi$ are the gravitational potentials, and we ignore tensor fluctuations.
We then obtain the action for the Goldstone boson $\pi$ up to second order: 
\bea
\label{fullpiaction}
&&\int d^4 x \sqrt{-g} \bigg[p_D+\dot{p}_D\pi+\half \ddot{p}_D\pi^2-\half(\rho_D+p_D)\left(2\Phi-2\dot{\pi}+4\Phi\dot{\pi}-\dot{\pi}^2+a^{-2}(\d \pi)^2\right)\\ \nonumber
&&\qquad\qquad\quad-(\dot{\rho}_D+\dot{p}_D) (\Phi-\dot{\pi}) \pi
+ 2M_2^4(t)(\Phi-\dot{\pi})^2\bigg] \ .
\eea
{At short distances, one can focus on the action of the Goldstone boson.} We can see that the kinetic part is given by
\be
S_{\rm kin.} = \int d^4 x \sqrt{-g} \left[ \half\left({\rho}_D + {p}_D + 4 M_2^4(t) \right) \dot \pi^2 - \half \left( {\rho}_D + {p}_D \right) a^{-2} \partial^2 \pi \right] \, ,
\ee
and thus the speed of sound is 
\be \label{speedofsounddeffull}
\css = \frac{ {\rho}_D + {p}_D}{ {\rho}_D + {p}_D + 4 M_2^4(t)} \, . 
\ee
The theory must be free of ghosts, which implies that the denominator has to be positive.
Therefore the speed of sound needs to have the same sign as $1+w$.
In particular, $w<-1$ implies $c_s^2<0$, which would produce gradient instabilities.
One can circumvent this instability by including the higher derivative terms proportional to $\hat{m}_{1,2}$, which scale as $k^4$ and give a stable dispersion relation at small scales~\cite{Creminelli:2006xe,Creminelli:2008wc}.
In order for the higher derivative terms not to be highly suppressed (which would make them irrelevant on cosmological scales), we need the speed of sound to be bound by $|c_s^2|<10^{-30}$, which means it is practically zero. {These considerations hold also when a careful analysis including the mixing with gravity is performed.} {Similar considerations are obtained by including the higher derivative operator proportional to $\bar m_1$~\cite{Creminelli:2006xe,Creminelli:2008wc}.}
In conclusion, {it {\it is} possible to have viable theories with $w<-1$, but they need to have} $c_s^2\rightarrow0$, which are called clustering dark energy or clustering quintessence. {We notice furthermore that in order to have a stable theory, we need to have $w\gtrsim -2$ if we use the operators in $\hat{m}_{1,2}$, or $\geq -1.17$ if we use the operator in $\bar m_1$~\cite{Creminelli:2006xe,Creminelli:2008wc}~\footnote{These lower limits will play essentially no role in our analysis, as the data constrain $w$ to be far from this boundary.}.}

The name ``clustering quintessence'' stems from the fact that the dark energy can cluster with the dark matter, and they jointly contribute to the gravitational potential. Hence the adiabatic mode (i.e. the perturbations of the total energy density, which source the gravitational potential) depends on both the dark matter and dark energy perturbations.
As a result, dark energy perturbations leave an imprint on biased tracers such as galaxies, which are the main interest in this work.
Therefore, next we wish to give a quick overview of how we derive the equations of motion for the adiabatic mode in the presence of clustering quintessence.

Before analyzing the equations for $\pi$, it is useful to write down the EFT equations for dark matter, which couples to dark energy through gravity~\cite{Lewandowski:2016yce}:
\begin{align}
	\label{fluid1}
&\dot{\delta}_m + \frac{1}{a} \partial_i ( (1 + \delta_m)v^i_m) =0 \, ,\\
	\label{fluid2}
&\partial_i \dot{v}^i_m + H \partial_i v^i_m + \frac{1}{a} \partial_i ( v^j_m \partial_j v^i_m ) + \frac{1}{a} \partial^2 \Phi = - \frac{1}{a} \partial_i \left( \frac{1}{\rho_m} \partial_j \tau^{ij} \right) \, ,
\end{align}
where $\delta_m$ and $v_m$ are the dark matter overdensity and velocity, $\dot{}=d/dt$ and $\tau^{ij}$ is the effective stress tensor.

Let us start analyzing the linear equations, and we will study the non-linear equations subsequently. The linear equation for $\pi$~\cite{Lewandowski:2016yce,Creminelli:2008wc, Creminelli:2009mu}, which we get from~\eqref{fullpiaction}, reads: 
\be \label{eom12}
\frac{1}{a^3}\frac{1 }{M_2^4}\frac{d}{dt } \left[ a^3 M_2^4 ( \dot \pi - \Phi) \right] = \frac{c_s^2}{1 - c_s^2} \frac{\partial^2 \pi}{a^2} \ .
\ee
This shows that, in the limit $c_s \to 0$, the RHS can be neglected.
We can, therefore, write $\dot{\pi}-\Phi \propto (a^3 M_2^4)^{-1}$, which is a decaying mode, assuming the speed of sound to be approximately constant.
In particular, we have $\d_i\dot{\pi}-\d_i\Phi=0$, and, using the linear-level Euler equation~\eqref{fluid2}, we get that $\sfrac{d}{dt}\left[a v_m^i +\d^i\pi\right] = 0$.
This means that on the growing adiabatic mode we have
\be\label{comove}
\d^i\pi=-a v_m^i \, ,
\ee
which implies that the two species are comoving.
This will eventually allow us to write a closed set of differential equations for the adiabatic mode, defined by $\delta_A = 2 M_{pl}^2 a^{-2} \partial^2 \Psi / \rhom$. The Poisson equation is~\cite{Creminelli:2008wc, Creminelli:2009mu, Gubitosi:2012hu}
\begin{align} \label{poisson123}
a^{-2} \partial^2 \Psi & = \frac{\rhom}{2 M_{pl}^2} \left(\delta_m + \frac{4 M_2^4 }{\rhom}( \dot \pi - \Phi) \right) \,.
\end{align}
Using the definition of the adiabatic mode, we find 
\be
\delta_A = \delta_m + \frac{4 a^3 M_2^4 }{\rho_{m,0}}( \dot \pi - \Phi)= \delta_m + \fr{(1+w)}{\css}\frac{ \rho_{D,0} }{\rho_{m,0}}a^{-3w}( \dot \pi - \Phi) \label{deltaalinear}\ . 
\ee
We can now take the derivative of the above equation and plug in the equation of motion for~$\pi$, Eq.~\eqref{eom12}, the solution for $\rho_D$, Eq.~\eqref{denssol}, and substitute the dark matter velocity for the spatial derivatives of $\pi$, Eq.~\eqref{comove}. We then get
\bea
\dot{\delta}_A &=& \dot{\delta}_m -\fr{1}{a} ( 1 + w) \fr{\rhodi}{ \rho_{m,0}} a^{-3w} \theta_m \label{deltaalineardot}\ . \\ \nonumber
&=&-\fr{1}{a}C(a)\theta_m,
\eea
where we have introduced the dark matter velocity divergence $\theta_m = \d_i v_m^i$ and we have defined
\be
C(a)=1+(1+w)\fr{\Omega_{D,0}}{ \Omega_{m,0}} a^{-3w} \, .
\ee

We now move on to the full non-linear equations of motion for the adiabatic mode, which is somewhat more technical.
We will just mention the main results and refer to \cite{Lewandowski:2016yce} ({see also~\cite{Fasiello:2016qpn}}) for more details.
First, we can easily see that the two species remain comoving at the non-linear level.
Using the equations of motion, one can show that $\delta g_u^{00} \propto c_s^2$ also at non-linear level.
Taking a spatial derivative, $\d_i \delta g_u^{00}$ = 0 in the limit $c_s^2 \to 0$, yields
 \begin{align}
 0 & = \partial_i \left( \dot \pi - \Phi - \half a^{-2} (\partial \pi)^2 \right) \\
 & = \frac{d}{dt} \left( a v_m^i + \partial_i \pi \right) + v_m^j \partial_j v_m^i - a^{-2} \partial_j \pi \partial_j \partial_i \pi \ . \label{velocityeq}
\end{align}
This is satisfied by simply using Eq.~\eqref{comove}, thus the two species are comoving also at non-linear level.
The full non-relativistic equation of motion for the dark energy field $\pi$ is given by
 \be\label{nonlineom}
 - \frac{2}{a^3} \partial_t \left( a^3 M_2^4 \,\deltagu \right) = M_2^4 \frac{c_s^2 }{1 - c_s^2} \frac{\partial^2 \pi}{a^2} - 2 a^{-2} \partial^2 \pi M_2^4 \deltagu \ , 
 \ee
where we used that $\d_i\delta g_u^{00} = 0$.
The full Poisson equation introduces non-linearities in the definition of the adiabatic mode, which reads
\be
 \delta_A = \delta_m - \fr{(1+w)}{2 \css}\frac{ \rho_{D,0} }{\rho_{m,0}}a^{-3w}\delta g_u^{00} \label{deltaanonlinear} \, . 
\ee

Now we can take a time derivative and obtain a non-linear continuity equation for the adiabatic mode.
The only difference is that we have to include the non-linear terms for $\dot{\delta}_m$ and we have an additional term in the equations of motion for $\pi$ on the right-hand side of Eq.~\eqref{nonlineom}.
We then get
\bea
 \dot{\delta}_A &=& -\fr{1}{a}C(a)\theta_m-\fr{1}{a}\d_i(\delta_m v_m^i)-\d^2\pi\fr{2 a M_2^4}{\rho_{m,0}}\delta g_u^{00}\\ \nonumber
 &=&-\fr{1}{a}C(a)\theta_m-\fr{1}{a}\d_i(\delta_m v_m^i)+\fr{1}{a}\theta_m(\delta_m-\delta_A)\\ \nonumber
 &=&-\fr{1}{a}C(a)\theta_m-\fr{1}{a}\d_i(\delta_A v_m^i), \label{deltaaeqnonlinear}\ 
\eea
where in the second line we use Eq.~\eqref{deltaanonlinear}, and in the last line we use $\d_i\delta_m=\d_i\delta_A$. 

Since the two species are comoving, $\theta_A = \theta_m$ and the Euler equation for the adiabatic mode is simply obtained by using the definition of the adiabatic mode in terms of the gravitational potential in Eq.~\eqref{fluid2}.
We finally get the governing equations for the clustering quintessence - dark matter system (without counterterms): 
 \begin{align}
 \dot \delta_A + \frac{1}{a} C(a) \theta_m & = - \frac{1}{a} \partial_i \left( \delta_A v_m^i \right) \label{nonlincont}\\
 \dot \theta_m + H \theta_m + \frac{3}{2} \frac{\Omega_{m,0} H_0^2}{a^2} \delta_A & = - \frac{1}{a} \partial_i \left( v_m^j \partial_j v_m^i \right) \label{nonlineuler} \, ,
 \end{align} 
where $ \rhom/(2 \mpl^2)= 3 \Omega_{m,0} \cH_0^2 a_0/ (2 a^3)$. {As explained in~\cite{Lewandowski:2016yce}, since clustering quintessence is comoving with dark matter, there is no isocurvature mode, and the counterterms are the same as for standard dark matter.} 
To solve the equations above perturbatively we transform into Fourier space, where they read ({still neglecting the counterterms}):
\bea\label{eq:master1}
&&a\delta'_{\vk}-f_{+}\theta_{\vk}=\fr{(2\pi)^{3}f_{+}}{C(a)}\iint \frac{d^3q_1}{(2\pi)^{3}}\frac{d^3q_2}{(2\pi)^{3}} \delta_{D}(\vk-\vq_1-\vq_2)\alpha(\vq_1,\vq_2)\theta_{\vq_1}\delta_{\vq_2},\\
&&a\theta'_{\vk}-f_{+}\theta_{\vk}- \frac{f_-}{f_{+}}(\theta_{\vk}-\delta_{\vk})=\fr{(2\pi)^{3}f_{+}}{C(a)}\iint \frac{d^3q_1}{(2\pi)^{3}}\frac{d^3q_2}{(2\pi)^{3}}\delta_{D}(\vk-\vq_1-\vq_2)\beta(\vq_1,\vq_2)\theta_{\vq_1}\theta_{\vq_2} \, ,
\label{eq:master2} 
\eea
and we drop the indices $m$ and $A$ from now on since we will only talk about the adiabatic mode.
We define $\delta=\delta_A$ and $\theta = -\fr{C}{f_+ a H}\d_i v^i$ for the rescaled velocity divergence such that $\delta^{(1)} =\theta^{(1)}$. Furthermore, we use the scale factor as time variable such that $'=d/da$ and defined the growth rates $f_{\pm} = \fr{d\,\text{ln\,}D_{\pm}}{d\,\text{ln}\,a}$ in terms of the growth factors, further discussed in Appendix \ref{appendixa}.
We will not use the commonly applied Einstein-de Sitter (EdS) approximation, where one approximates the time dependence of a perturbation by powers of the growth factor, for instance $\delta^{(n)}_{\vk}(a) \stackrel{\rm EdS}{\propto} D^n(a)\delta^{(n)}_{\vk}(a_i)$, for some intital time $a_i$.
Instead, we will use the exact time dependence solution discussed below. As we will see later, the EdS approximation significantly biases the determination of the cosmological parameters in the presence of clustering quintessence.

Eqs.~\eqref{eq:master1}-\eqref{eq:master2} are slightly different from the dark matter equations in the presence of smooth dark energy with $c_s^2=1$, i.e. $w$CDM.
In fact, in the limit {$(1+w)\rightarrow 0$, with $\Omega_{D,0}=$const}, we recover, {at large distances where we can neglect the higher derivative terms,} the equations of motion for the matter overdensity in $\Lambda$CDM. 
This difference in the equations of motion between the two models results in a modified definition of the time functions that appear in the exact time solutions for $\delta$ and $\theta$. 
Exact solutions for the adiabatic mode $\delta$ in the presence of clustering quintessence have been previously studied in \cite{Lewandowski:2016yce,Sefusatti:2011cm,Fasiello:2016qpn}.
The time-dependent integral kernel solutions in Fourier space are given by~\cite{Lewandowski:2016yce}
\bea\label{kernels1}
K_\lambda^{(1)}(\vq_1,a)&=& 1 \, ,\\
\label{kernels2}
K_\lambda^{(2)}(\vq_1,\vq_2,a)&=&\alpha_s(\vq_1,\vq_2)\mG^{\lambda}_{1}(a)+\beta(\vq_1,\vq_2)\mG^{\lambda}_{2}(a) \, ,\\
\label{kernels3}
K_\lambda^{(3)}(\vq_1,\vq_2,\vq_3,a)&=&\alpha^{\sigma}(\vq_1,\vq_2,\vq_3)\mU^{\lambda}_{\sigma}(a) +\beta^{\sigma}(\vq_1,\vq_2,\vq_3)\mV^{\lambda}_{\sigma2}(a) + \gamma^{\sigma}(\vq_1,\vq_2,\vq_3)\mV^{\lambda}_{\sigma1}(a) \, ,
\eea
where repeated $\sigma \in \{1,2\}$ are summed over and $\lambda \in \{\delta,\theta\}$.
The explicit time functions are defined in Appendix \ref{appendixa}, and the momentum functions in Appendix \ref{appendixb}.
The kernels in Eqs.~\eqref{kernels1}-\eqref{kernels3}, and in the following sections are defined by
\bea\label{kerneldef}
X^{(n)}(\vk,a) =\int\frac{d^3q_1}{(2\pi)^{3}}...\frac{d^3q_n}{(2\pi)^{3}}(2\pi)^{3}\delta_{D}(\vk-\vq_1-...-\vq_n) \ K_{X}^{(n)}(\vq_1,...,\vq_n,a)\ \delta^{(1)}_{\vq_1}(a)...\delta^{(1)}_{\vq_n}(a) \, ,
\eea
where $X$ may for instance stand for $\delta$ or $\theta$.
In the next section, we will see how the solution with exact time dependence for clustering quintessence leaves an imprint in the bias expansion of biased tracers such as galaxies.

\subsection{Perturbative expansions of $\delta_h$ and $\theta_h$}
To find the bias expansion for the galaxy overdensity $\delta_h$ following the exact time dependence solution of the adiabatic mode, we can follow a procedure similar to \cite{Donath:2020abv}. Ref.~\cite{Fujita:2020xtd} has also recently derived the same results, using a different approach.
Here equations will change with respect to~\cite{Donath:2020abv}, as a consequence of the modified equations of motion for $\delta_A$, relative to the equations for the dark matter solutions in $w$CDM.
As has been previously studied in~\cite{Senatore:2014eva}, the bias expansion for $\delta_h$ is given by
\bea\label{eq:euler_bias_raw}
&&\delta_h(\vec x,a)\simeq \int^a \fr{da'}{a'}\; \left[ c_{\delta}(a,a')\; :\delta(\xfl,a'): \right.\\ \nonumber
&&\quad+ c_{\delta^2}(a,a')\; :\delta(\xfl,a')^2: + c_{s^2}(a,a')\; :s^2(\xfl,a'):\\\nonumber
&&\quad+ c_{\delta^3}(a,a')\; :\delta(\xfl,a')^3 : + c_{\delta s^2}(a,a')\; : \delta(\xfl,a') s^2(\xfl,a'):+ c_{\psi}(a,a')\; :\psi(\xfl,a'):\\ \nonumber
&&\quad+ c_{st}(a,a') \; :st(\xfl,a'):+ c_{ s^3}(a,a')\; :s^3(\xfl,a'):\\\nonumber
&&\quad+ c_{\epsilon}(a,a')\;\epsilon(\xfl,a')\\ \nonumber
&&\quad+ c_{\epsilon\delta}(a,a') \;:\epsilon(\xfl,a')\delta(\xfl,a'):+ c_{\epsilon s}(a,a') \;:\epsilon s(\xfl,a'):+ c_{\epsilon t }(a,a') \;:\epsilon t(\xfl,a'):\\\ \nonumber
&&\quad+ c_{\epsilon^2\delta}(a,a') \;:\epsilon(\xfl,a')^2\delta(\xfl,a'):
+ c_{\epsilon\delta^2}(a,a') \;:\epsilon(\xfl,a')\delta(\xfl,a')^2:+ c_{\epsilon s^2}(a,a') \;:\epsilon(\xfl,a')s^2(\xfl,a'): \\ \nonumber
&&\quad+ c_{\epsilon s \delta}(a,a') \;:\epsilon s(\xfl,a')\delta(\xfl,a'):+ c_{\epsilon t \delta}(a,a') \;:\epsilon t(\xfl,a')\delta(\xfl,a'):\\ \nonumber
&&\left.\quad+ c_{\d^2\delta}(a,a')\; \;\frac{\d^2_{x_{\rm fl}}}{\km^2}\delta(\xfl,a')+\dots\ \right] \, ,
\eea
where we include all possible operators\footnote{The notation $:\mathcal{O}:$ means that the operator $\mathcal{O}$ is subtracted of its vacuum expectation value, i.e. $:\mathcal{O}:=\mathcal{O}-\langle\mathcal{O}\rangle$.} allowed by the equivalence principle, including stochastic contributions and higher derivative terms.
Their definitions are found in Appendix~\ref{appendixb}. {As for the dark matter equations, since clustering quintessence is comoving with dark matter, there is no isocurvature mode, and the bias expansion depends on the same fields as for the dark-matter-only universe~\cite{Lewandowski:2016yce}.}
{The time-kernels, such as $c_\delta(a,a')$, that account for the time non-locality, can be formally integrated over $a'$ after the perturbative solutions are substituted in}. All operators (which are explicitly given in Appendix~\ref{appendixb}) are evaluated along the fluid line-element:
\bea\label{xfl}
\xfl(\vec x, a,a') = \vec x - \int_{a'}^{a} \fr{da''}{a''^2H(a'')} \vec{v}(a'',\vec x_{\rm fl}(\vec x, a,a'')).
\eea
This results in Taylor expansions in the fields around $\vx$ given by \bea\label{eq:xfl_expansion}
&&\delta(\xfl(a,a'),a')=\delta(\vec x,a')-\d_i\delta(x,a')\int_{a'}^a 	\fr{da''}{a''^2H(a'')} \; v^i(\vec x,a'')\\ \nonumber
&&\quad \quad\quad\quad\quad\quad+\frac{1}{2}\d_i\d_j \delta(x,a')\int_{a'}^a 	\fr{da''}{a''^2H(a'')}\; v^i(\vec x,a'')\int_{a'}^a 	\fr{da'''}{a'''^2H(a''')}\;v^j(\vec x,a''')\\ \nonumber
&&\quad \quad\quad \quad\quad \quad+\d_i\delta(x,a')\int_{a'}^a 	\fr{da''}{a''^2H(a'')}\; \d_j v^i(\vec x,a'') \int^a_{a''} 	\fr{da'''}{a'''^2H(a''')}\;v^j(\vec x,a''')+\ldots\ .
\eea 
It turns out that even in the presence of clustering quintessence, once we perturbatively expand the overdensity and velocity, the time integrals in Eq.~\eqref{eq:xfl_expansion} can be done analytically and the solutions are given in terms of the time functions and kernels that appear in Eqs.~\eqref{kernels1}-\eqref{kernels3}.
This is explicitly derived in Appendix~\ref{appendixc}.
{Then, as mentioned before, after perturbatively expanding the fields, the time integrals in Eq.~\eqref{eq:euler_bias_raw} are formally done, and result in the definition of coefficients such as}
\be
\label{eq:coefs}
c_{\delta,1}(a)= \int^a \frac{da'}{a'}c_{\delta}(a,a')\frac{D_{+}(a')}{D_{+}(a)} , \qquad c_{\delta^2,1}(a)=\int^a \frac{da'}{a'}c_{\delta^2}(a,a')\frac{D_{+}(a')^2}{D_{+}(a)^2} ,\quad \ldots\ .
\ee
For a complete list see Appendix \ref{appendixb}.
After this procedure, the resulting halo overdensity can then be written as a sum of functions of time multiplied by functions of momentum.
As was shown in~\cite{Donath:2020abv}, some of the momentum functions are degenerate and can all be expressed in terms of the basis $\{\mathbb{I},\alpha,\beta,\alpha_1,\alpha_2,\beta_1,\beta_2,\gamma_1,\gamma_2\}$, which are the kernels that appear in Eqs.~\eqref{kernels1}-\eqref{kernels3}.
This is true in $w$CDM as well as the clustering quintessence case, because the momentum functions are the same in both cases, and only the time functions change.
We can therefore write
\bea\label{eq:euler_bias_k}
K_{\delta_h}^{(1)}(\vq_1,a)&=& c_{\delta,1}(a) \, ,\\ \nonumber 
K_{\delta_h}^{(2)}(\vq_1,\vq_2,a)&=&c_{\mathbb{I},(2)}(a)+c_{\alpha,(2)}(a)\alpha(\vq_1,\vq_2)+c_{\beta,(2)}(a)\beta(\vq_1,\vq_2) \, , \\ \nonumber
K_{\delta_h}^{(3)}(\vq_1,\vq_2,\vq_3,a)&=&c_{\alpha_\sigma,(3)}(a)\alpha^\sigma(\vq_1,\vq_2,\vq_3) +c_{\beta_\sigma,(3)}(a)\beta^\sigma(\vq_1,\vq_2,\vq_3)+c_{\gamma_\sigma,(3)}(a)\gamma^\sigma(\vq_1,\vq_2,\vq_3) \\ \nonumber
&&+c_{\alpha,(3)}(a)\alpha(\vq_1,\vq_2)+c_{\beta,(3)}(a)\beta(\vq_1,\vq_2) +c_{\mathbb{I},(3)}(a) \, ,
\eea
where in the last expression a sum is implied over $\sigma \in \{1,2\}$. The main reason that the time coefficients $c_i$ change, relative to $w$CDM, is because the integrals from the flow terms that stem from the Taylor expansion of Eq.~\eqref{eq:xfl_expansion} now have an additional dependence on $C(a)$ (for a comparison see Appendix \ref{appendixc}).
The coefficients in Eq.~\eqref{eq:euler_bias_k} are explicitly defined in Appendix~\ref{appendixb}.
For more details on the derivation of the halo overdensity kernels, see~\cite{Donath:2020abv}.

From here we can proceed in a very similar fashion to \cite{Donath:2020abv}.
We reduce the number of coefficients by looking for degeneracies in the time coefficients.
Luckily all the identities from~\cite{Donath:2020abv} still hold in a slightly more general form.
The main difference here is that we define the calculable function $\mG = \mG_1^{\delta}+ \mG_2^{\delta}$, {with $\mG_i^{\delta}$ defined in App. \ref{appendixa}}, which for $w$CDM is $\mG\stackrel{w\text{CDM}}{=}1$. The identities then read
\bea
\label{eq:degen}
&&c_{\alpha,(2)}+c_{\beta,(2)} = \mG\, c_{\delta,1} \, , \\ \nonumber
&&c_{\alpha,(3)}+c_{\beta,(3)} = 2\,\mG\, c_{\mathbb{I},(2)} \, , \\ \nonumber
&&c_{\beta_2,(3)}+\mG\, c_{\alpha,(2)}-c_{\alpha_1,(3)}= \fr{\mG^2}{2}c_{\delta,1} \, , \\ \nonumber
&&c_{\alpha_1,(3)}+c_{\alpha_2,(3)}=c_{\gamma_1,(3)}+c_{\gamma_2,(3)} \, , \\ \nonumber
&&c_{\beta_1,(3)}+c_{\beta_2,(3)}+c_{\gamma_1,(3)}+c_{\gamma_2,(3)} =	\fr{\mG^2}{2}c_{\delta,1} \, , \\ \nonumber
&&c_{\gamma_1,(3)}+c_{\beta_1,(3)}=\left(\fr{3}{14} +Y(a)\right)c_{\delta,1} \, , 
\eea
where in the limit $\mG\stackrel{w\text{CDM}}{=}1$ we recover the identities from \cite{Donath:2020abv}.
$Y(a)$ is defined by
\bea
Y(a)=-\fr{3}{14}+\mV^{\delta}_{11}(a)+\mV^{\delta}_{12}(a).
\eea 
However, it is useful to define
\bea\label{Ynice}
\tilde Y(a)=-\fr{3}{14}\mG(a)^2+\mV^{\delta}_{11}(a)+\mV^{\delta}_{12}(a),
\eea 
so that, taking limits, we have $\tilde Y(a) \stackrel{w\text{CDM}}{=}Y(a)\stackrel{\text{EdS}}{=}0$.
We can then write the final halo overdensity ({see also~\cite{Fujita:2020xtd}}):
\bea \label{eq:delta_h_CoI_t}
\delta_h(\vec k,a)&=& c_{\delta,1}(a) \; \Big( \mathbb{C}^{(1)}_{\delta}(\vec k,a)+\mG(a)\mathbb{C}^{(2)}_{\delta}(\vec k,a)+\mG(a)^2\mathbb{C}^{(3)}_{\delta}(\vec k,a) +\tilde Y(a)\mathbb{C}^{(3)}_{Y}(\vec k,a)\Big)\\ \nonumber
&+& c_{\alpha,(2)}(a) \; \Big(\mathbb{C}^{(2)}_{\alpha}(\vec k,a) +\mG(a)\mathbb{C}^{(3)}_{\alpha_1}(\vec k,a)\Big)\\ \nonumber &+&c_{\mathbb{I},(2)}(a) \;\Big(\mathbb{C}^{(2)}_{\mathbb{I}}(\vec k,a)+2\,\mG(a)\mathbb{C}^{(3)}_{\beta}(\vec k,a)\Big) \\ \nonumber
&+& c_{\beta_1,(3)}(a) \; \mathbb{C}^{(3)}_{\beta_1}(\vec k,a)+ c_{\gamma_2,(3)}(a) \; \mathbb{C}^{(3)}_{\gamma_2}(\vec k,a)\\ \nonumber
&+&c_{\alpha,(3)}(a) \; \mathbb{C}^{(3)}_{\alpha}(\vec k,a)+c_{\mathbb{I},(3)}(a) \mathbb{C}^{(3)}_{\mathbb{I}}(\vec k,a) \, ,
\eea
where we can see that no new $\mathbb{C}_i$ operators have to be included compared to the exact $w$CDM case or EdS approximated case.
The $\mathbb{C}_i$ are defined in the same way as in~\cite{Donath:2020abv} and are explicitly given in Appendix~\ref{appendixb}.
Similarly to what happens when we use the exact time dependence for smooth dark energy {and $\Lambda$CDM}, we see that there are additional calculable time dependencies in the final bias expansion for the galaxy overdensity. However, there are no new bias coefficients. We can take two interesting limits to see how the above expansion generalizes previous models.
In the $\mG \rightarrow 1$ limit, we obtain the galaxy overdensity in $w$CDM with exact time dependence.
Furthermore, in the limit where we use the EdS approximation, the time functions in Eqs.~\eqref{kernels1}-\eqref{kernels3} become independent of~$a$ and with a value so that $\mG \rightarrow 1$ and $	\tilde Y\rightarrow 0$~\footnote{Note that in the presence of clustering quintessence the EdS approximation does not only rely on $\Omega_m/f_+^2\approx1$ but also on $C(a)\approx1$ which is needed to cancel the time dependence in the continuity equation~\eqref{eq:master1}. Therefore, from Eq.~\eqref{Gint} one can see that in the EdS approximation one takes the limit $\mG\rightarrow1$.}. Eq.~\eqref{eq:delta_h_CoI_t} can then simply be linearly transformed into the BoD basis from \cite{Angulo:2015eqa}, therefore the space spanned by the kernels in Eq.~\eqref{eq:delta_h_CoI_t} is the same as the one spanned by the BoD basis from \cite{Angulo:2015eqa} (for a transformation see \cite{Donath:2020abv}). For illustration, we plot in Fig.~\ref{fig:YG} the values of $\tilde Y$ and $\mG$ as functions of the redshift $z=1/(1+a)$ and $w$.

\begin{figure}[h]
\centering
\includegraphics[width=0.99\textwidth]{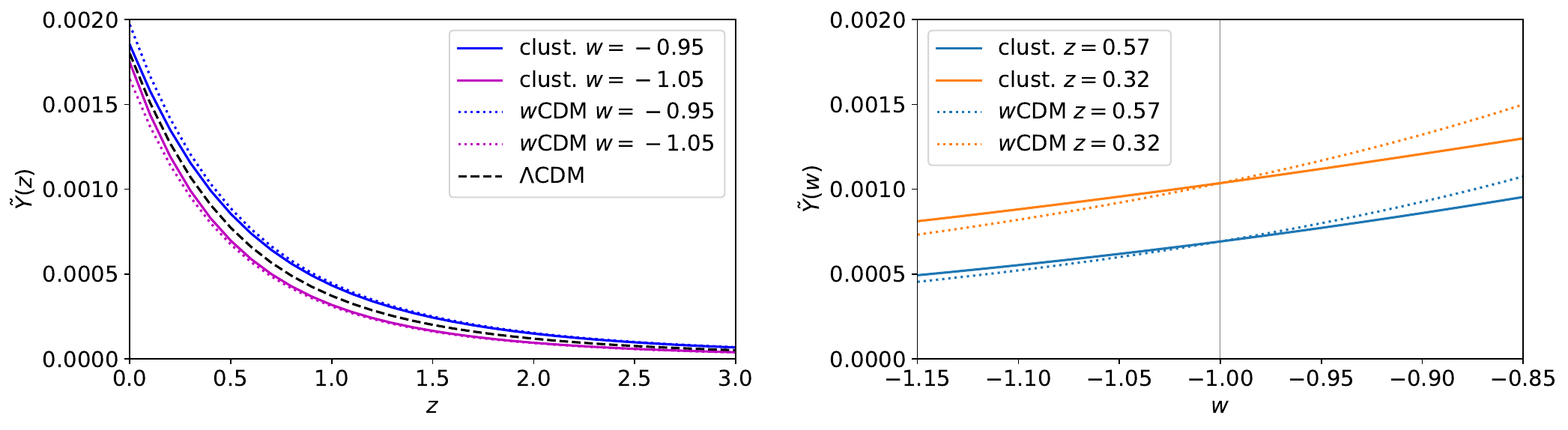}
\includegraphics[width=0.99\textwidth]{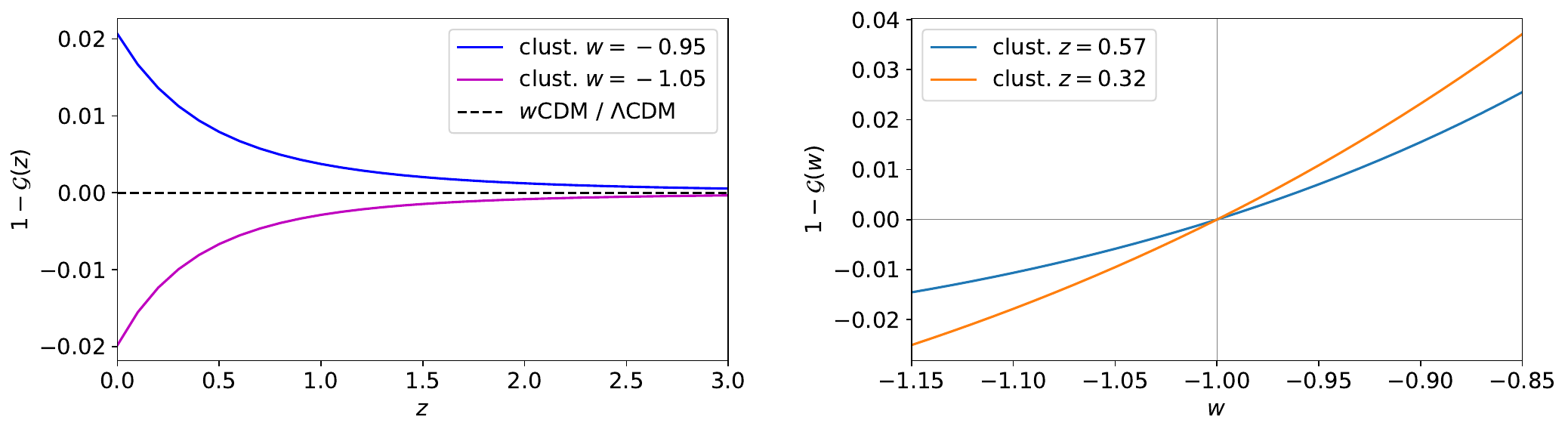}
\caption{\small $\tilde Y$ and $\mG$ as a function of redshift $z$ and quintessence equation of state parameter $w$. 
We show $\Lambda$CDM and $w$CDM cases for comparison. 
Notice that, as we argued earlier, for $w <-1$ we need $c_s^2 \rightarrow 0$ and thus for $c_s^2 =1$, i.e. $w$CDM, $w<-1$ is not allowed in the EFT of dark energy. We, nevertheless, plot it here for illustration.}

\label{fig:YG}
\end{figure}

In a last step, we write the expansion for $\theta_h$, which appears in the redshift space expansion.
For the velocity divergence, there is no bias \cite{Perko:2016puo}, up to higher derivative terms.
We can thus model the velocity divergence as a species of biased tracer.
Specifically, we obtain the velocity divergence by plugging in the following choice of functions into Eq.~\eqref{eq:delta_h_CoI_t}:
\bea
c_{\delta,1}^{\theta}(a)&=&1\\ \nonumber
c_{\mathbb{I},(2)}^{\theta}(a)&=& c_{\mathbb{I},(3)}^{\theta}(a)=c_{\alpha,(3)}^{\theta}(a)=0\\ \nonumber
c_{\alpha,(2)}^{\theta}(a)&=&\mG^{\theta}_{1}(a) \\ \nonumber 
c_{\beta_1,(3)}^{\theta}(a) &=&\mV^{\theta}_{12}(a)\; \\ \nonumber 
c_{\gamma_2,(3)}^{\theta}(a)&=&\mV^{\theta}_{21}(a).
\eea
The counterterms will take the exact same form as for $w$CDM \cite{Lewandowski:2016yce,Donath:2020abv}.
We will now transform into redshift space and compute the power spectrum.

\subsection{Galaxy Power spectrum in redshift space}
As the next step, we wish to compute the full galaxy power spectrum in redshift space, which we will later use to fit the data.
As shown in~\cite{Donath:2020abv}, the EdS approximation has no influence on the transformation into redshift space~\footnote{Of course since the halo overdensity in redshift space depends on $\delta_h$ and $\theta_h$, the exact time dependence has an impact, just not on the transformation itself.}.
This means we can proceed in the same way as described in~\cite{Perko:2016puo}.
The galaxy overdensity kernels in redshift space in terms of the real space quantities $\delta_h$ and $\theta_h$ are given by (without counterterms)
\begin{align}\label{eq:redshift_kernels}
K_{\delta_{h,r}}^{(1)}(\q_1,a) & = K_{\delta_h}^{(1)}(\q_1,a) +f_+ \, \mu_1^2 K_{\theta_h}^{(1)}(\q_1,a) = b_1 + f_+ \, \mu_1^2,\\ \nonumber
K_{\delta_{h,r}}^{(2)}(\q_1,\q_2,\mu,a) & = K_{\delta_h}^{(2)}(\q_1,\q_2,a) + f_+ \, \mu_{12}^2 K_{\theta_h}^{(2)}(\q_1,\q_2,a) \\ \nonumber
&+ \, \frac{1}{2}f_+ \, \mu \, q \left[ \frac{\mu_2}{q_2}K_{\theta_h}^{(1)}(\q_2,a) K_{\delta_{h,r}}^{(1)}(\q_1,a) + \text{perm.} \right],\\ \nonumber
K_{\delta_{h,r}}^{(3)}(\q_1,\q_2,\q_3,\mu,a) & = K_{\delta_h}^{(3)}(\q_1,\q_2,\q_3,a) + f_+ \, \mu_{123}^2 K_{\theta_h}^{(3)}(\q_1,\q_2,\q_3,a) \nonumber\\ \nonumber
& + \frac{1}{3}f_+ \, \mu \, q \left[ \frac{\mu_3}{q_3} K_{\theta_h}^{(1)}(\q_3,a) 	K_{\delta_{h,r}}^{(2)}(\q_1,\q_2,\mu_{123},a) \right. \\ \nonumber
& \left. \quad + \frac{\mu_{23}}{q_{23}}K_{\theta_h}^{(2)}(\q_2,\q_3,a)	K_{\delta_{h,r}}^{(1)}(\q_1,a)+ \text{cyc.} \right],
\end{align}
where $\delta_{h,r}$ is the halo overdensity in redshift space. Using $\hat{z}$ as the line of sight unit vector, we have defined $\mu= \q \cdot \hat{z}/q$, with $\q = \q_1 + \dots +\q_n$, and $\mu_{i_1\ldots i_n} = \q_{i_1\ldots i_n} \cdot \hat{z}/q_{i_1\ldots i_n}$, $\q_{i_1 \dots i_m}=\q_{i_1} + \dots +\q_{i_m}$. 	
As we mentioned previously, the counterterms and stochastic terms that come from real and redshift space (see \cite{Perko:2016puo,Donath:2020abv} for a discussion) do not change in the presence of clustering quintessence. Therefore, the final expression for the galaxy power spectrum in redshift space, including the counterterms, reads
\begin{align}\label{eq:powerspectrum}
P_{g}(k,\mu,a) & = 	K_{\delta_{h,r}}^{(1)}(\mu,a)^2 P_{11}(k,a) \nonumber \\
& + 2 \int \frac{d^3q}{(2\pi)^3}\; 	K_{\delta_{h,r}}^{(2)}(\q,\k-\q,\mu,a)^2 P_{11}(|\k-\q|,a)P_{11}(q,a) \\ \nonumber
&+ 6 	K_{\delta_{h,r}}^{(1)}(\mu,a) P_{11}(k,a) \int\, \frac{d^3 q}{(2\pi)^3}\; 	K_{\delta_{h,r}}^{(3)}(\q,-\q,\k,\mu,a) P_{11}(q,a)\nonumber \\ \nonumber
& + 2 	K_{\delta_{h,r}}^{(1)}(\mu,a) P_{11}(k,a)\left( c_\text{ct}\frac{k^2}{{ k^2_\textsc{m}}} + c_{r,1}\mu^2 \frac{k^2}{k^2_\textsc{m}} + c_{r,2}\mu^4 \frac{k^2}{k^2_\textsc{m}} \right) \\ \nonumber
& + \frac{1}{{n}_g}\left( c_{\epsilon,0}+c_{\epsilon,1}\frac{k^2}{k_\textsc{m}^2} + c_{\epsilon,2} f_+\mu^2 \frac{k^2}{k_\textsc{m}^2} \right),
\end{align}
where $P_{11}(k,a)$ is the time-dependent linear power spectrum for the adiabatic mode, $k_\textsc{m}\lesssim \knl$ is the comoving wavenumber which controls the bias derivative expansion, and ${n}_g$ is the background galaxy number density.
In the first line, we have the linear power spectrum in redshift space.
In the second and third line, we have the $P_{13}$ and $P_{22}$ contributions of the loop and in the fourth and fifth line we have the counterterms and stochastic terms, respectively.

Finally, the power spectrum is IR-resummed following~\cite{Senatore:2014vja,Senatore:2017pbn,Lewandowski:2018ywf,DAmico:2020kxu}. Since quintessence is comoving with dark matter, the equations for the IR-resummation only change by a shift $P_{11}(k,a)\rightarrow \mG(a)^2 P_{11}(k,a)$ \footnote{The additional factors of $\mG$ come from the integral over the velocity, when expressing the displacement field in terms of the overdensity\bea
		s^{(1)j}(a) = \int^a \fr{da'}{a'^2 H(a')}v^{(1)j}(a').
\eea
This integral, which is explicitly computed for the flow terms in Eq.~\eqref{appendix-flow}, results in $s^{(1)j}(a)= - \mG(a) \fr{\d^j}{\d^2}\delta^{(1)}$.}.
We then apply corrections to take into account the Alcock-Pacszynski effect~\cite{Alcock:1979mp}, window functions~\cite{Beutler:2018vpe}, and fiber collisions~\cite{Hahn:2016kiy}.

\begin{figure}[h]
\centering
\includegraphics[width=0.99\textwidth]{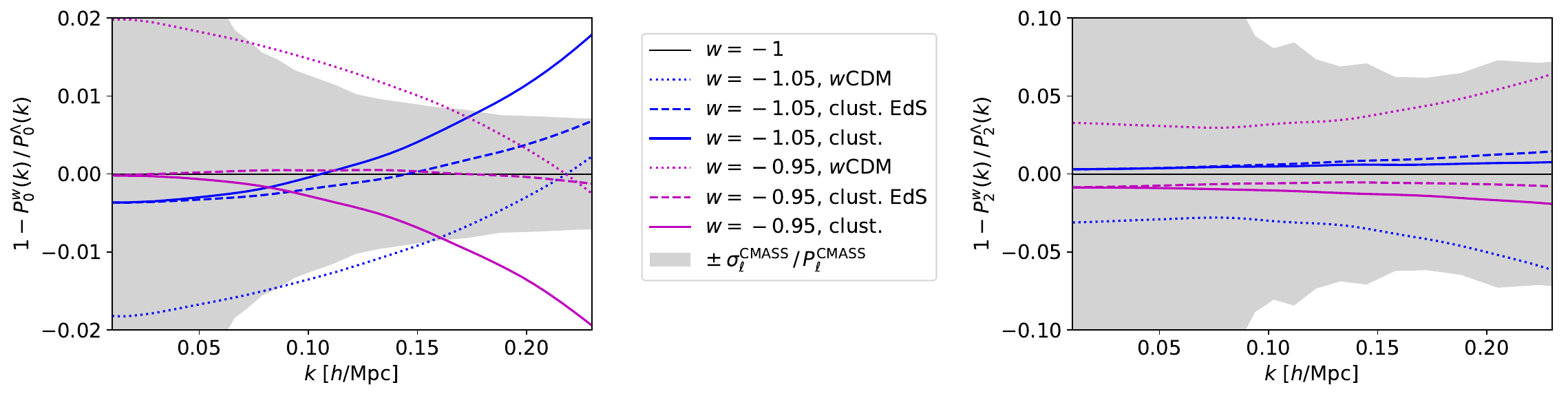}
\caption{\small One-loop galaxy power spectrum multipole ratio of $w$CDM or clustering quintessence, with $w=-0.95$ and $w=-1.05$, to $\Lambda$CDM, at $z=0.57$. 
We show for clustering quintessence the evaluation with and without the EdS approximation. 
The evaluation in $\Lambda$CDM and $w$CDM are with exact time dependence. 
The EFT parameters are the same for all evaluations, with values as the best fit of BOSS on $\Lambda$CDM. 
The BOSS CMASS error bars are depicted for comparison. 
Notice that, as we argued earlier, for $w <-1$ we need $c_s^2 \rightarrow 0$ and thus for $c_s^2 =1$, i.e. $w$CDM, $w<-1$ is not allowed in the EFT of dark energy. We, nevertheless, plot it here for illustration.}
\label{fig:pk}
\end{figure}

In Fig.~\ref{fig:pk}, we show the difference between the one-loop galaxy power spectrum multipoles $\ell=0,2$ evaluated in different cosmologies: $\Lambda$CDM, $w$CDM and clustering quintessence, for $w=-0.95$ and $w=-1.05$. 
We also show the difference between the evaluation with and without the EdS approximation for clustering quintessence. 
It is apparent that the difference between $w$CDM and clustering quintessence is important with respect to the BOSS error bars. 
The difference between the evaluation with and without the EdS approximation for clustering quintessence is clearly important, especially in the monopole. 
Given how large the differences in the power spectrum are, we expect to see differences at the level of the posteriors of the cosmological and EFT parameters.

\section{LSS data analysis}\label{sec:analysis}

In this section, after calibrating the scale cut of the theory against simulations, we present the results from fitting clustering and smooth quintessence to the BOSS FS, and its combinations with BAO, SN and CMB measurements. 

\paragraph{Likelihood and priors}
The theory prediction is given by the galaxy power spectrum in redshift space at one-loop order in the EFTofLSS, eq.~\eqref{eq:powerspectrum}.
Its evaluation is performed using \code{PyBird}~\cite{DAmico:2020kxu}, and we sample from a Gaussian likelihood.
The kernels $K_{\delta_{h,r}}^{(1)}$, $K_{\delta_{h,r}}^{(2)}$, $K_{\delta_{h,r}}^{(3)}$ depend on 4 biases: $b_1$, $b_2$, $b_3$, $b_4$.
In our analysis, we use the monopole and quadrupole of the galaxy power spectrum.
We vary the cosmological parameters $\omega_{cdm}, h$, $\ln (10^{10}A_s)$, $n_s$, $w$, on which we impose no priors, and $\omega_b$ with a Gaussian prior motivated from Big Bang Nucleosynthesis (BBN) with standard deviation $\sigma_{\omega_b, {\rm BBN}} = 0.00036$~\cite{Mossa:2020gjc}.
For the simulations, we will center the prior on the truth, while on BOSS data, we will use $\omega_{b,{\rm BBN}} = 0.02233$~\cite{Mossa:2020gjc}. 
When analyzing BOSS data~\cite{Alam:2016hwk} (alone and in combination with other datasets) we fix the neutrinos to minimal mass, $0.06 \, {\rm eV}$, as done in the Planck analysis~\cite{Planck:2018vyg}.
As for the EFT parameters, we define the linear combinations $c_2 = (b_2 + b_4) / \sqrt{2}$, $c_4 = (b_2 - b_4) / \sqrt{2}$, and we set $c_4 = 0$ since $b_2$, $b_4$ are almost completely anticorrelated.
Then, we define the two combinations $c_{\epsilon, \rm mono} = c_{\epsilon,1} + f c_{\epsilon,2} / 3$, $c_{\epsilon, \rm quad} = 2 f c_{\epsilon,2} / 3$.
We put a Gaussian prior of mean 0 and standard deviation 2, $\mathcal{N}(0,2)$, on $b_3$, $c_\text{ct}$, $c_{\epsilon,0}$, $c_{\epsilon, \rm mono}$, $c_{\epsilon, \rm quad}$, and a Gaussian prior of mean 0 and standard deviation 8, $\mathcal{N}(0,8)$, on the redshift-space counterterm $c_{r,1}$.
We fix $c_{r,2}=0$ since it is exactly degenerate with $c_{r,1}$ when only analyzing the monopole and quadrupole.
As explained in~\cite{DAmico:2019fhj,DAmico:2020kxu}, we analytically marginalize over $b_3$, $c_{\rm ct}$, $c_{\epsilon, 0}$, $c_{\epsilon, \rm mono}$, $c_{\epsilon, \rm quad}$, $c_{r,1}$ as they appear linearly in the power spectrum and therefore quadratically in the likelihood.
Finally, the linear bias $b_1$ has a flat prior $[0, 4]$, and $c_2$ has a flat prior $[-4,4]$, which play no role.

\subsection{Tests against simulations}\label{sec:scalecut}

To assess the theory-systematic error of the FS analysis, we fit the power spectrum multipoles measured from large-volume N-body simulations on clustering quintessence with a BBN prior.
We consider two independent realizations of the BOSS `lettered' challenge simulations, which are boxes of side length $2.5 \, {\rm Gpc}/h$, described in e.g.~\cite{DAmico:2019fhj}.
The first realization is made of four boxes, labelled A, B, F, and G, populated by four different Halo Occupation Distribution~(HOD) models, of which we analyze the snapshot at $z=0.56$.
The second realization, labelled D, is populated by another HOD model, of which we analyze the snapshot at $z=0.5$.
Using one box, we can measure for each cosmological parameter the theory-systematic error as the distance in the 1D posterior of the $1\sigma$ region to the truth of the simulation.
Therefore, the theory-systematic error is zero if the truth lies within the $1\sigma$ region.
For A, B, F, and G, which are correlated, we average the posteriors for the cosmological parameters, that we label ABFG.
Moreover, we can combine ABFG with D, as they are independent realizations, allowing us to measure the theory error using a volume about $14$ times the one of BOSS data.
To do so, we combine for each cosmological parameter the 1D posterior of the shift of the mean with respect to the truth, as the product of two Gaussian distributions.
The distance of the $1\sigma$ region to zero in each resulting 1D posterior gives a measure of the theory-systematic error for the combination ABFG+D. 
For each cosmological parameter, the error bar obtained on ABFG+D represents the smallest theory-systematic error {which we can measure}, which is between $0.3\cdot \sigma_{\rm data}$ and $0.5 \cdot \sigma_{\rm data}$, where $\sigma_{\rm data}$ is the error bar obtained by fitting BOSS data.

\begin{table}[h]
\footnotesize
\centering
\begin{tabular}{l|c|c|c|c|c|c} 
 \hline 
 		& $\omega_{cdm}$ & $h$ & $\ln (10^{10} A_s)$ & $n_s$ & $w$ & $\Omega_m$				\\ 
& $\sigma_{\rm stat} | \sigma_{\rm sys}$ & $\sigma_{\rm stat} | \sigma_{\rm sys}$ & $\sigma_{\rm stat} | \sigma_{\rm sys}$ & $\sigma_{\rm stat} | \sigma_{\rm sys}$ & $\sigma_{\rm stat} | \sigma_{\rm sys}$ & $\sigma_{\rm stat} | \sigma_{\rm sys}$ \\ \hline 
ABFG 		& $0.007 | 0.000 $ & $0.027 | 0.000$ & $0.11 | 0.04$ & $0.044 | 0.000$ & 	$0.139 | 0.000$ & $0.021 | 0.000$	 \\ \hline 
D			& $0.006 | 0.000 $ & $0.018 | 0.000$ & $0.11 | 0.04$ & $0.039 | 0.000$ &	$0.093 | 0.000$ & $0.014 | 0.000$	 \\ \hline 
ABFG+D		& $0.005 | 0.000 $ & $0.015 | 0.000$ & $0.08 | 0.07$ & $0.029 | 0.000$ &	$0.077 | 0.000$ & $0.012 | 0.000$	 \\ \hline 
\end{tabular}
 \caption{\small 68\%-confidence intervals $\sigma_{\rm stat}$ and theory-systematic errors $\sigma_{\rm sys}$ obtained fitting clustering quintessence to the lettered challenge simulations with a BBN prior. }
 \label{tab:challenge}
\end{table}

\begin{figure}[h]
\centering
\includegraphics[width=0.49\textwidth]{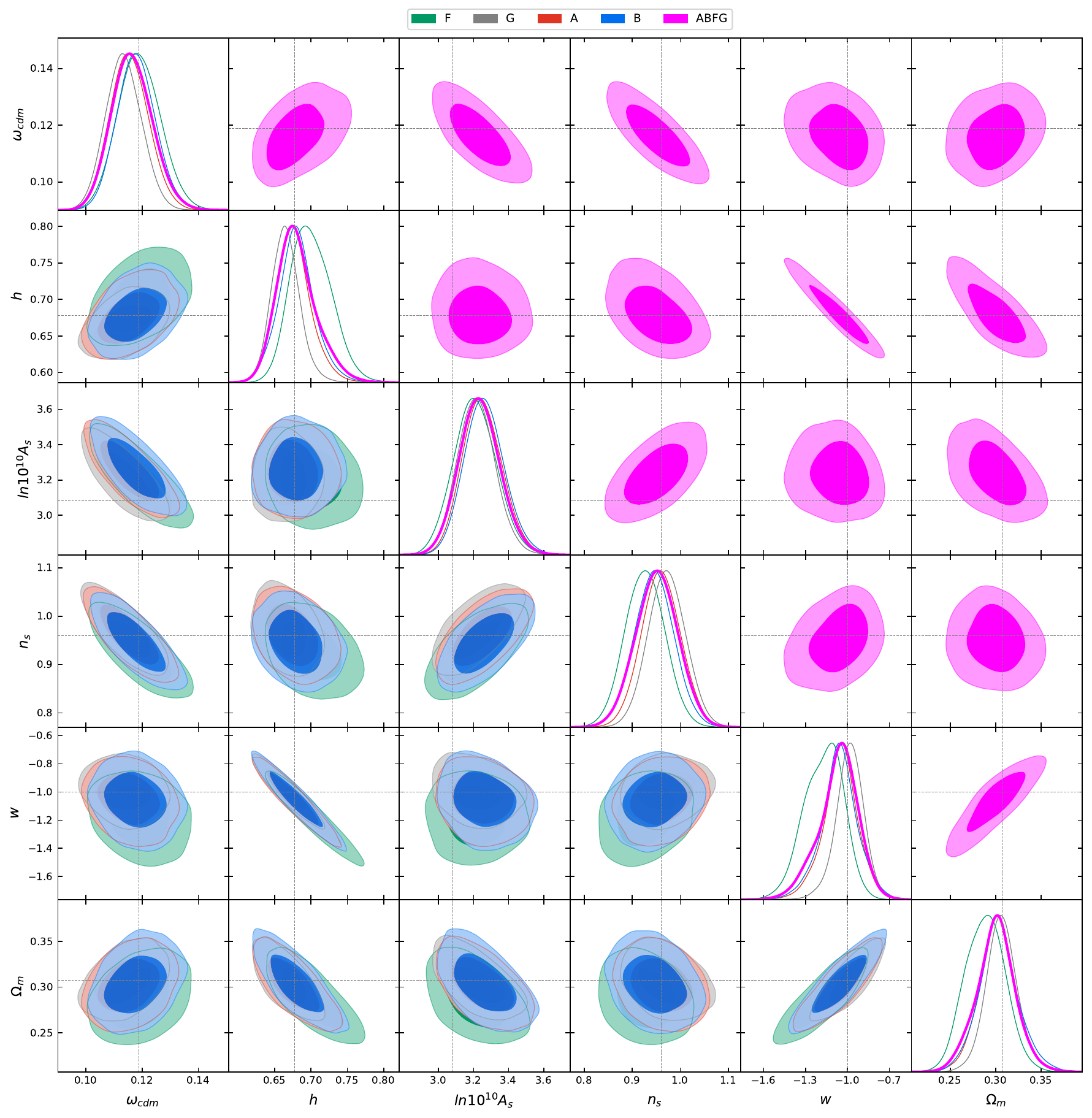}
\includegraphics[width=0.49\textwidth]{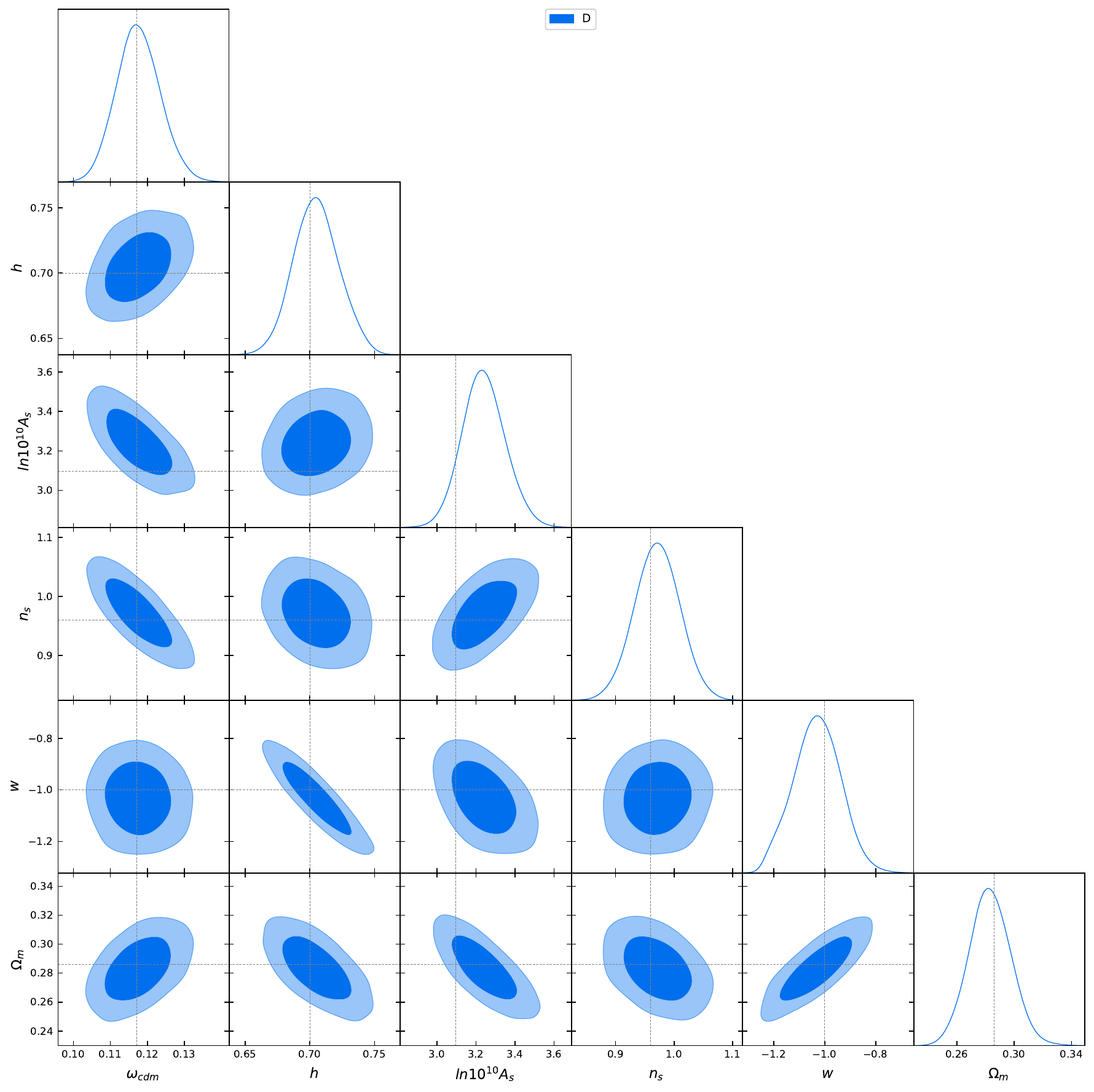}
\caption{\small Triangle plots obtained by fitting clustering quintessence to the lettered challenge simulations with a BBN prior.
The dashed lines represent the truth of the simulations.}
\label{fig:challenge}
\end{figure}

In Fig.~\ref{fig:challenge} and Tab.~\ref{tab:challenge}, we show the results obtained by fitting the lettered challenge simulations at scale cut $\kmax = 0.23 \hinvMpc$. 
We find for all cosmological parameters zero theory-systematic error, with the exception of $\ln (10^{10} A_s)$, where we find a marginal theory-systematic error of {$0.07$, which is $\sim 0.4\cdot \sigma_{\rm data}$~\footnote{Given the number of cosmological parameters, we find the likelihood of such a large value of one cosmological parameter to be sufficiently high, so that we do not include this in the systematic error budget.}.}
These results show that we can confidently fit the data up to $\kmax = 0.23 \hinvMpc$ on our high redshift ($z_{\rm eff}=0.57$) sample CMASS. 
For LOWZ sample at $z_{\rm eff}=0.32$, we rescale the scale cut as in~\cite{DAmico:2019fhj} and fit up to $\kmax = 0.2 \hinvMpc$. 

\subsection{LSS constraints}\label{sec:results}

In Fig.~\ref{fig:qcdm} and Tab.~\ref{tab:qcdm}, we show the results obtained by fitting BOSS FS+BAO, and in combination with BAO measurements from 6DF/MGS and eBOSS, and with Pantheon SN, on clustering quintessence with a BBN prior. 
We see that all cosmological parameters can be measured (we do not quote $\omega_b$ since it is dominated by the BBN prior we impose).
For all analyses performed, $w$ is consistent with $-1$ at $\lesssim 1\sigma$.

\begin{table}
\footnotesize
\centering
\begin{tabular}{|l|c|c||c|c||c|c|} 
 \hline 
		& \multicolumn{2}{c||}{BOSS} 					& \multicolumn{2}{c||}{BOSS+6DF/MGS+eBOSS} 		& \multicolumn{2}{c|}{BOSS+6DF/MGS+eBOSS+SN} \\ \hline
 		 & best-fit & mean$\pm\sigma$ 					& best-fit & mean$\pm\sigma$ 						& best-fit & mean$\pm\sigma$ \\ \hline 
$\omega_{cdm }$ &$0.1271$ & $0.1346_{-0.016}^{+0.011}$ &$0.1188$ & $0.122_{-0.0099}^{+0.0083}$ 			&$0.1196$ & $0.1234_{-0.01}^{+0.008}$\\ 
$H_0$ &$66.75$ & $67.58_{-3.5}^{+2.7}$					&$66.99$ & $67.35_{-2.3}^{+2}$					&$67.97$ & $68.72_{-1.6}^{+1.4}$\\ 
$\ln (10^{10}A_{s })$ &$2.733$ & $2.64_{-0.17}^{+0.16}$ &$2.837$ & $2.79_{-0.16}^{+0.14}$			&$2.848$ & $2.806_{-0.16}^{+0.15}$\\ 
$n_{s }$& $0.9103$ & $0.8884_{-0.059}^{+0.072}$ 		&$0.9406$ & $0.9416_{-0.051}^{+0.053}$ 				&$0.972$ & $0.9335_{-0.05}^{+0.054}$\\ 
$w$ &$-0.878$ & $-0.8666_{-0.15}^{+0.17}$				&$-0.9212$ & $-0.9358_{-0.092}^{+0.11}$ 			&$-0.9928$ & $-1.011_{-0.048}^{+0.053}$ \\ \hline
$\Omega_m$ &$0.337$ & $0.3456_{-0.027}^{+0.03}$ 		&$0.3166$ & $0.3197_{-0.015}^{+0.017}$ 			&$0.3083$ & $0.3099_{-0.011}^{+0.012}$\\ 
$\sigma_8$ &$0.684$ & $0.6675_{-0.067}^{+0.061}$ 		&$0.7043$ & $0.7034_{-0.057}^{+0.047}$				&$0.7371$ & $0.7285_{-0.049}^{+0.043}$ \\ 
\hline 
 \end{tabular} \\
 \caption{Results obtained by fitting clustering quintessence to BOSS in combination with other late-time probes with a BBN prior.}
 \label{tab:qcdm}
\end{table}

\begin{figure}[h!]
\centering
\includegraphics[width=0.99\textwidth]{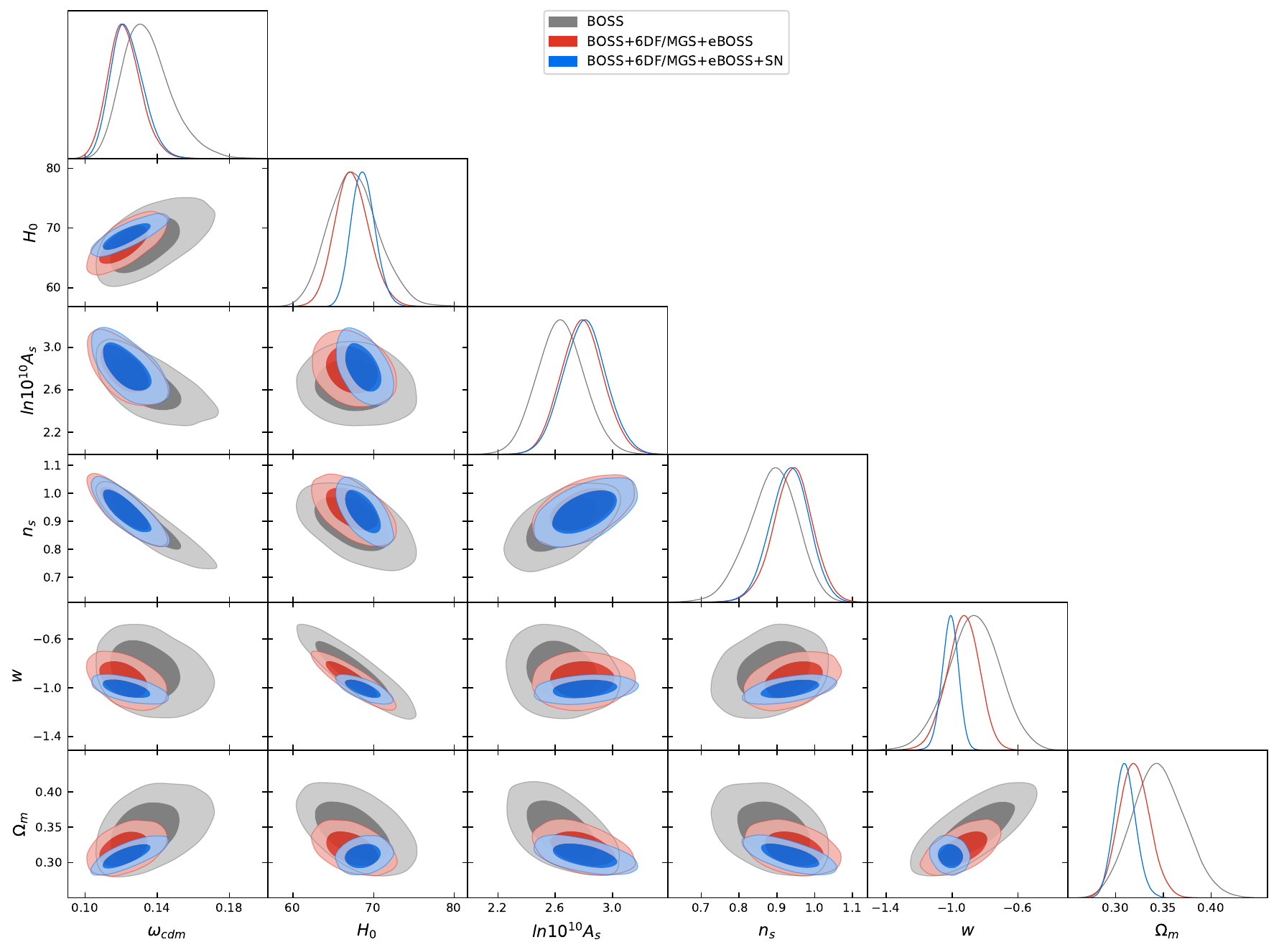}
\caption{\small Triangle plots obtained by fitting clustering quintessence to BOSS in combination with other late-time probes with a BBN prior.}
\label{fig:qcdm}
\end{figure}

\paragraph{Physical considerations} 
We now discuss why all cosmological parameters can be measured by analyzing the FS using the EFTofLSS, and how the addition of the SN measurements helps to obtain better constraints. Let us start with the contribution from the BAO information.
The two angles corresponding to the BAO components perpendicular and parallel to the line of sight are given by: 
\begin{equation}
\theta_{\rm LSS, \perp} \simeq \frac{r_d(z_{\rm CMB})}{D_A(z_{\rm LSS})} \, \qquad \theta_{\rm LSS, \parallel} \simeq \frac{r_d(z_{\rm CMB})}{c z_{\rm LSS}/H(z_{\rm LSS})} \, .
\end{equation}
Here $r_d(z_{\rm CMB})$ is the sound horizon at the end of the baryon-drag epoch $z_{\rm CMB}$, and $D_A(z_{\rm LSS})$ and $H(z_{\rm LSS})$ are the angular diameter distance and the Hubble parameter at the effective redshift of the survey $z_{\rm LSS}$. As discussed in e.g.~\cite{DAmico:2019fhj,DAmico:2020kxu}, these angles carry information about $h,\Omega_m$ and $w$. The dependence on parameters is the same as in $w$CDM, as the angles only depend on the background geometry~\cite{DAmico:2020kxu}:
\bea\nn
&&\theta_{\rm LSS, \, \parallel}(z_{{\rm Ly}\alpha}) \sim \Omega_m^{0.17} h^{0.42} |w|^{-0.11},\qquad
\theta_{\rm LSS, \, \perp}(z_{{\rm Ly}\alpha}) \sim \Omega_m^{0.01} h^{0.48} |w|^{-0.19} \ ,\\ \nn
&&\theta_{\rm LSS, \, \parallel}(z_{\rm CMASS}) \sim \Omega_m^{-0.02} h^{0.49} |w|^{-0.25},\qquad
\theta_{\rm LSS, \, \perp}(z_{\rm CMASS}) \sim \Omega_m^{-0.12} h^{0.53} |w|^{-0.17} \ ,\\ \nn
&&\theta_{\rm LSS, \, \parallel}(z_{\rm LOWZ}) \sim \Omega_m^{-0.10} h^{0.52} |w|^{-0.21}, \qquad
\theta_{\rm LSS, \, \perp}(z_{\rm LOWZ}) \sim \Omega_m^{-0.16} h^{0.54} |w|^{-0.12} \ ,\\ 
&&\theta_{\rm LSS, \, V}(z_{\rm 6dF}) \sim \Omega_m^{-0.19} h^{0.55} |w|^{-0.07},\qquad
\theta_{\rm LSS, \, V}(z_{\rm MGS}) \sim \Omega_m^{-0.18} h^{0.55} |w|^{-0.09} \ .
\eea
where $z_{{\rm Ly}\alpha}=2.35$, $z_{\rm CMASS}=0.57$, $z_{\rm LOWZ}=0.32$ and $z_{\rm 6dF/MGS}=0.106$. $\theta_{\rm LSS, \, V}$ is a combination of $\theta_{\rm LSS, \, \perp}$ and $\theta_{\rm LSS, \, \parallel}$ (see e.g.~\cite{DAmico:2020kxu}). 
{The dependences on the cosmological parameters above and in the rest of this section are obtained expanding around a fiducial cosmology ($\Omega_m=0.3$, $h=0.7$, $w=-1$).}
Furthermore, the relative amplitude of the BAO wiggles with respect to the smooth part instead gives a measurement of $\sim \Omega_m h^2$ (though the amplitude is not part of the standard BAO analysis). Clearly, at least in principle, this information allows for a determination of $w,\Omega_m$ and $h$. Notice however that the measurements for $w$ and $\Omega_m$ are strongly degenerate when using solely the BAO information from CMASS and LOWZ, and the breaking of the degeneracy by measuring both $\theta_{\rm LSS, \perp}$ and $\theta_{\rm LSS, \parallel}$ is mild, insufficient to get strong constraints~\cite{DAmico:2020kxu}. Of course, the situation is greatly ameliorated by the addition of the information from 6dF/MGS and eBOSS, but it is also ameliorated by the inclusion of the FS analysis.

In fact, the FS contains information not only through the BAO signal, but also by its shape and amplitude~\cite{DAmico:2019fhj}. The shape depends on the equality scale, and therefore on $\Omega_m h^2$. The amplitude and the anisotropy of the FS can be roughly summarized by the fact that the monopole and quadrupole mainly depend on the combinations $b_1(z)^2 D_+(z)^2 A_s^{(\kmax)}$ and $b_1(z) f_+(z) D_+(z)^2 A_s^{(\kmax)}$.
Here $A_s^{(\kmax)}$ is the amplitude of the linearly evolved power spectrum at the maximum wavenumber of our analysis, $A_s^{(\kmax)}\sim \left(k/k_0\right)^{n_s-1} \left(k_{\rm eq}/\kmax\right)^2 A_s$, with $k_{\rm eq}$ being the wavenumber that re-enters the horizon at equality and $k_0$ the pivot scale. $D_+$ and $f_+$ are respectively the growth factor and growth rate of the growing adiabatic mode. $\kmax$ is the maximum wavenumber of our analysis, which is where the signal to noise is dominated. Given that there are two redshifts in BOSS, this clearly offers a way to measure both $A_s$ and $n_s$, together with $b_1(z_{\rm CMASS})$ and $b_1(z_{\rm LOWZ})$. In this way, all cosmological parameters are, at least in principle, measured. However, we should keep in mind that the FS offers an independent measurement for each wavenumber, therefore, by combining the information from several $k$'s, further information on $w$ and $\Omega_m$ is obtained. In fact, just by looking at the dependence at linear level of the monopole and quadrupole at $z_{\rm CMASS}$ and $z_{\rm LOWZ}$, one can see that on top of $b_1$ and $A_s$, one can measure the combination $\left.\frac{f(z_{\rm CMASS}) D(z_{\rm CMASS})}{f(z_{\rm LOWZ}) D(z_{\rm LOWZ})}\right|_{\rm clust.}$, which, around the fiducial cosmology, goes as $\sim \Omega_m^{-0.12} |w|^{0.44}$. This can be seen by using the fitting functions for $D_+$ and $f_+$ as a function of $\Omega_m$ and $w$ given in~\cite{Sefusatti:2011cm}, which read:
\bea
&&\frac{D_+(a)}{a} = \frac{5}{2} \Omega_m(a) \left[ \Omega_m(a)^{4/7} + \frac{3}{2} \Omega_m(a) + \left( \frac{1}{70} - \frac{1+w}{4} \right) \Omega_D(a) \left( 1 + \frac{\Omega_m(a)}{2}\right)\right]^{-1}\ , \\\nn
&&f_+(a) = C(a) \left[ \Omega_m(a)^{4/7} + \left( \frac{1}{70} - \frac{1+w}{4} \right) \Omega_D(a) \left( 1 + \frac{\Omega_m(a)}{2}\right)\right]\ ,
\eea
where $C(a) = 1 + (1+w) \Omega_D(a)/\Omega_m(a)$. 
This is to be contrasted with the same ratio for the case of a smooth dark energy component, namely $w$CDM, around the same cosmology: $\left.\frac{f(z_{\rm CMASS}) D(z_{\rm CMASS})}{f(z_{\rm LOWZ}) D(z_{\rm LOWZ})}\right|_{w{\rm CDM}}\sim \Omega_m^{-0.12} |w|^{0.006}$. We can see that the {change in the} dependence on $w$ going from LOWZ to CMASS is stronger in the case of clustering quintessence compared to $w$CDM, physically originating from the fact that clustering quintessence contributes to the clustering. The mild degeneracy present for $w$CDM between $\Omega_m$ and $w$ is thus less pronounced in clustering quintessence when jointly fitting LOWZ and CMASS. Furthermore, these measurements give different correlations between $\Omega_m$ and $w$ with respect to the ones in $\theta_{\rm LSS}$, thus further breaking the degeneracies. This can be seen in Fig.~\ref{fig:qcdm_vs_wcdm}, where we compare the posteriors obtained fitting BOSS FS+BAO on clustering quintessence and $w$CDM. To summarize, $\Omega_m$, $h$, $w$, $A_s$, $n_s$ and $b_1$ can be measured from the BAO angles in combination with the broadband signal.

By looking at the same Fig.~\ref{fig:qcdm_vs_wcdm}, one can also see that in $w$CDM there is a large degeneracy in lowering $w$ and lowering $A_s$. This can be explained by the fact that, in $w$CDM with $w<-1$ (which, we remind, is physically inconsistent at the quantum level but can still be analyzed as a model), matter domination lasts longer, so that structures grow more and therefore the power spectrum is left unchanged by lowering $A_s$. In clustering quintessence, this degeneracy is broken by the fact that the adiabatic mode receives a contribution from clustering quintessence proportional to $1+w$. This can be see from solving the linear equations, which, at early times, give~(see e.g.~\cite{Lewandowski:2016yce}, eq. (4.15)):
\be
\delta_A(a_{\rm early})=\left(1+\frac{(1+w)}{1-3w}\frac{\Omega_{D,0}}{\Omega_{m,0}}\left(\frac{a_{\rm early}}{a_{0}}\right)^{-3w}\right) \delta_m(a_{\rm early})\ ,
\ee
with $a_0$ the present epoch and $a_{\rm early}$ a time early on during matter domination.
This effect acts in a direction contrary to the extra growth that one gets from the extension of the epoch of matter domination for $1+w<0$, in practice bounding the degeneracy between $w$ and $A_s$.

\begin{figure}[h]
\centering
\includegraphics[width=0.99\textwidth]{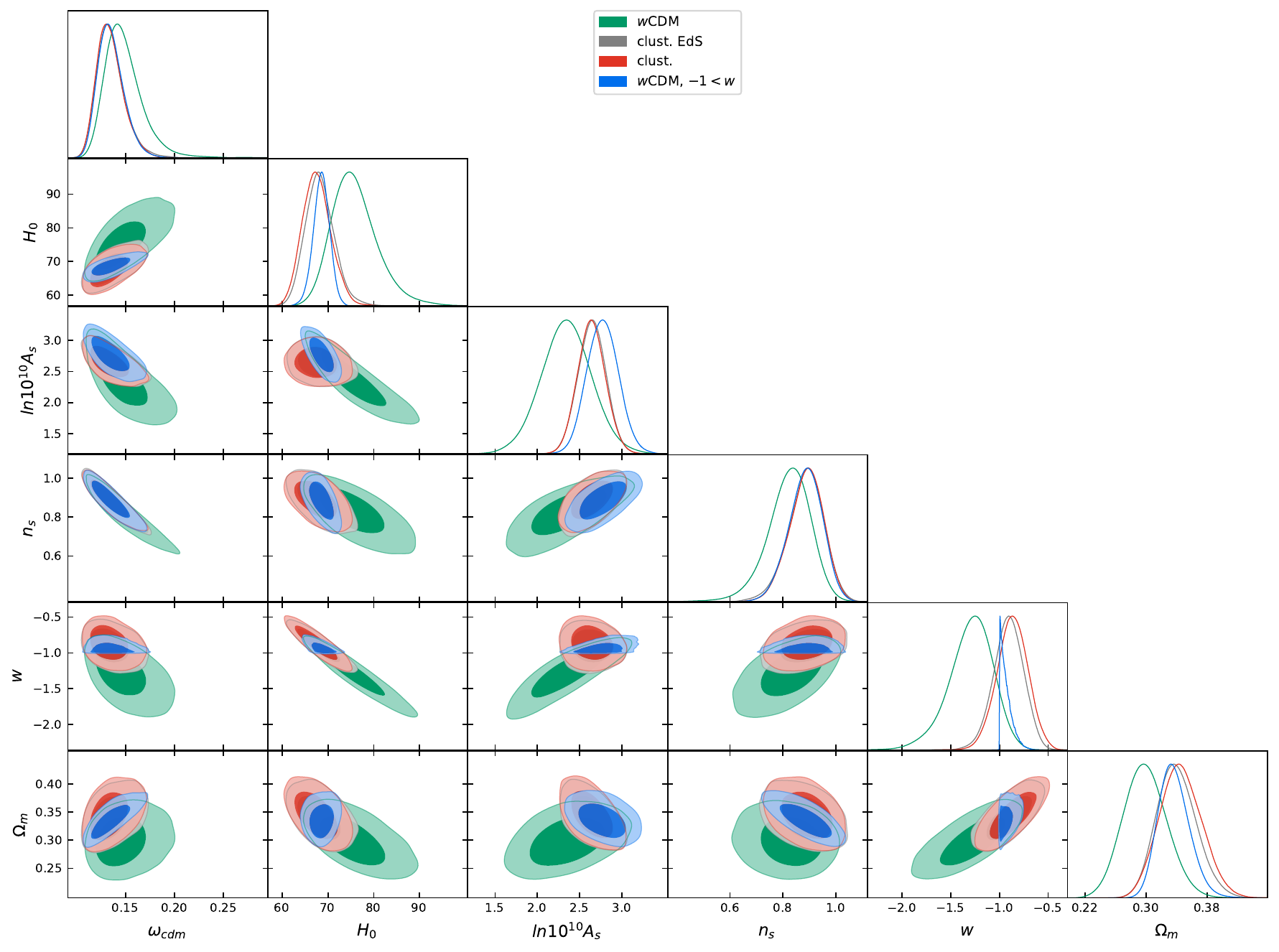}
\caption{\small Triangle plots obtained by fitting clustering quintessence to BOSS with a BBN prior, with or without the EdS approximation. For comparison, we show $w$CDM fit to BOSS with a BBN prior, with and without physical prior $w \geq 1$. }
\label{fig:qcdm_vs_wcdm}
\end{figure}

Note that this discussion gives only rough estimates of the parameter dependence of the FS. 
In practice, there is no separation between the broadband and the other sources of information within the FS analysis as all the signal is analyzed up to the chosen scale cut. 
In particular, the loop provides additional information. 
For example, the growth function enters as $D_+^4$ in the loop, providing yet another parametric dependence on $w$. 
In Fig.~\ref{fig:qcdm_vs_wcdm}, we also show the posteriors obtained on clustering quintessence with the EdS approximation~\footnote{In appendix~\ref{appendixd} we show the full posterior including the nuisance parameters}.
The difference with the posteriors obtained with exact time dependence is clearly visible: most notably, about $0.2 \, \sigma$ for $H_0$ and $\Omega_m$, and $0.3 \, \sigma$ for $w$.
At the level of the power spectrum in Fig.~\ref{fig:pk}, the difference is somewhat larger in terms of error bars, but we should remember that in that figure the EFT parameters are fixed. 
In particular, the large deviation that can be seen in the monopole of Fig.~\ref{fig:pk} can be partially absorbed below the error bars with a small offset in the shot noise $c_{\epsilon, 0}/ n_g$ of $\sim 0.1$. 
The difference we see between the EdS evaluation and the exact-time one can be traced to the time functions, as for example $\mG^2$, in some loop terms when evaluated with exact time dependence: $\mG(z_{\rm LOWZ})^2 \sim |w|^{0.42}$ and $\mG(z_{\rm CMASS})^2 \sim |w|^{0.27}$}.
Because of this, the EdS approximation leads to noticeable shifts in the posteriors for clustering quintessence.
The shifts are of the same order as the theory-systematic error we find in sec.~(3.1), so it may look like we can neglect them.
However, we would then introduce an additional systematic error on the parameters.
Contrary to the uncertainty from next orders in perturbation theory, the exact time dependence can be easily computed, with the same computational cost and without adding new nuisance parameters. Therefore, we prefer to use the exact time dependence.

Finally, the distance-redshift relation of SN data from Pantheon brings evidently more constraints. Approximately, the line degeneracy of the luminosity distance $D_L = (1+z)^2 D_A$ is $D_L(z = 0.25) = \Omega_m^{-0.05} |w|^{0.1}$, which further helps break the degeneracy between $\Omega_m$ and $w$ when fitting jointly with the FS and BAO.

\subsection{CMB+LSS constraints}

\begin{figure}[h!]
\centering
\vspace{10pt}
\includegraphics[width=0.99\textwidth]{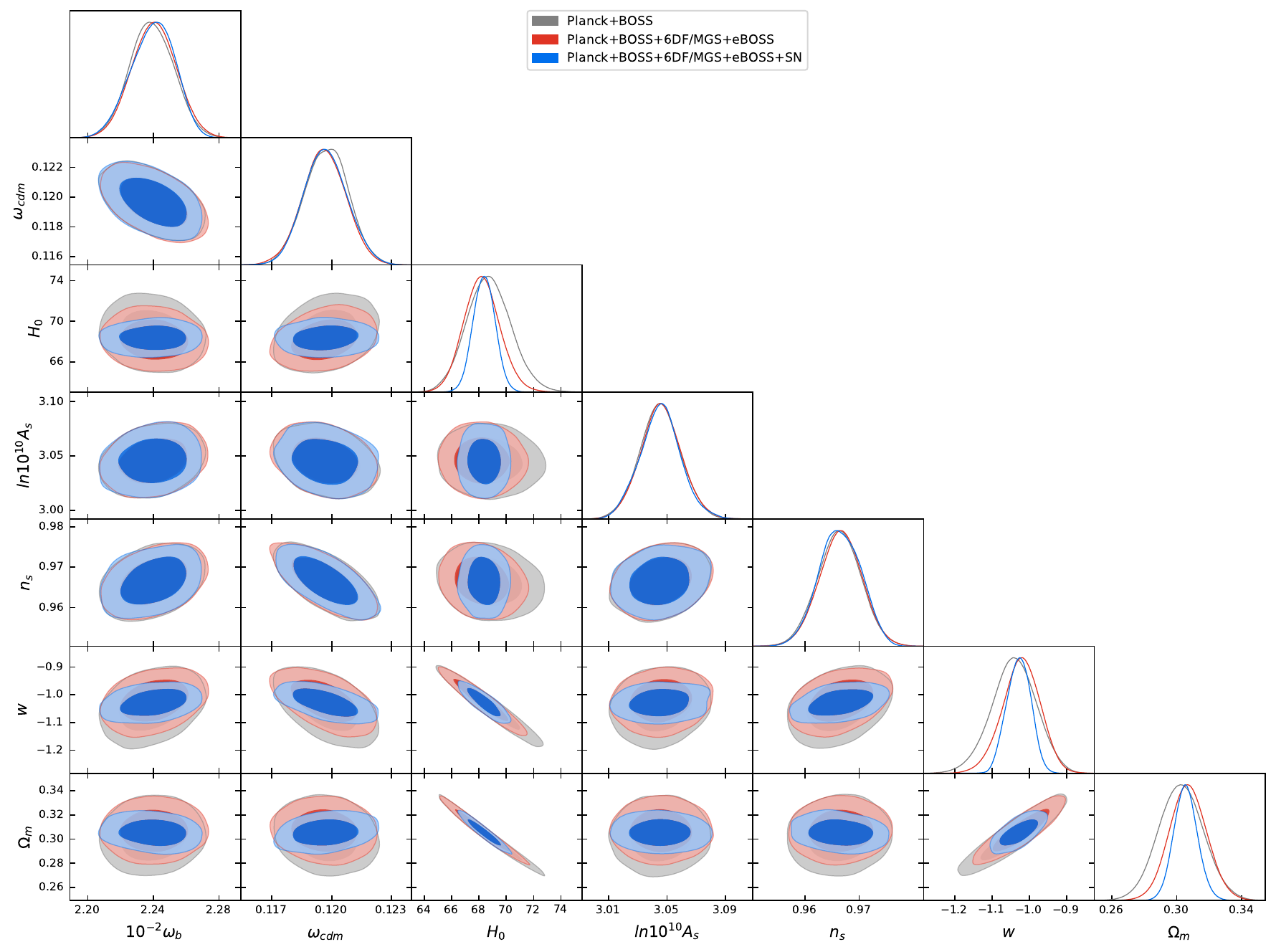}
\caption{\small Triangle plots obtained by fitting clustering quintessence to Planck and BOSS in combination with other late-time probes.}
\label{fig:qcmb}
\end{figure}

In Fig.~\ref{fig:qcmb} and Tab.~\ref{tab:qcmb}, we show the results obtained fitting clustering quintessence with Planck data in combination with BOSS FS+BAO, BAO measurements from 6DF/MGS and eBOSS and with Pantheon SN. 

\begin{table}[h]
\footnotesize
\centering
\begin{tabular}{|l|c|c||c|c||c|c|} 
 \hline 
	Planck +	& \multicolumn{2}{c||}{BOSS} 					& \multicolumn{2}{c||}{BOSS+6DF/MGS+eBOSS} 		& \multicolumn{2}{c|}{BOSS+6DF/MGS+eBOSS+SN} \\ \hline
 		 & best-fit & mean$\pm\sigma$ 					& best-fit & mean$\pm\sigma$ 						& best-fit & mean$\pm\sigma$ \\ \hline 
$100~\omega_{b }$ &$2.238$ & $2.239_{-0.014}^{+0.014}$ 	&$2.241$ & $2.24_{-0.014}^{+0.014}$ 				&$2.239$ & $2.24_{-0.012}^{+0.015}$ \\
$\omega_{cdm }$ &$0.12$ & $0.1197_{-0.0011}^{+0.0011}$ 	&$0.1196$ & $0.1196_{-0.0011}^{+0.0011}$			&$0.1197$ & $0.1197_{-0.0012}^{+0.00099}$\\ 
$H_0$ &$68.17$ & $68.74_{-1.7}^{+1.6}$				&$68.15$ & $68.22_{-1.4}^{+1.2}$ 					&$68.54$ & $68.38_{-0.84}^{+0.78}$\\ 
$\ln (10^{10}A_{s })$ &$3.041$ & $3.045_{-0.015}^{+0.014}$ 	&$3.047$ & $3.046_{-0.014}^{+0.014}$				&$3.049$ & $3.046_{-0.014}^{+0.014}$\\ 
$n_{s }$ &$0.9648$ & $0.9663_{-0.0039}^{+0.0042}$			&$0.9681$ & $0.9666_{-0.0039}^{+0.0041}$			&$0.9647$ & $0.9665_{-0.0036}^{+0.0042}$ \\
$\tau_{reio }$ &$0.05156$ & $0.05487_{-0.0078}^{+0.007}$	&$0.05677$ & $0.05534_{-0.0075}^{+0.007}$			&$0.05732$ & $0.05507_{-0.0071}^{+0.0072}$ \\
$w$ &$-1.027$ & $-1.041_{-0.058}^{+0.064}$				&$-1.019$ & $-1.022_{-0.047}^{+0.056}$	 			&$-1.034$ & $-1.028_{-0.030}^{+0.037}$ \\ \hline
$\Omega_m$ &$0.3079$ & $0.3027_{-0.014}^{+0.014}$ 		&$0.3073$ & $0.307_{-0.012}^{+0.012}$ 			&$0.304$ & $0.3055_{-0.0073}^{+0.0074}$\\ 
$\sigma_8$ &$0.8162$ & $0.8213_{-0.018}^{+0.017}$		&$0.8164$ & $0.8162_{-0.017}^{+0.015}$				&$0.8204$ & $0.8179_{-0.012}^{+0.0097}$ \\ 
$100~\theta_{s }$ &$1.042$ & $1.042_{-0.0003}^{+0.0003}$	&$1.042$ & $1.042_{-0.00028}^{+0.0003}$		&$1.042$ & $1.042_{-0.00033}^{+0.00026}$ \\
\hline 
 \end{tabular} \\
 \caption{Results obtained by fitting clustering quintessence to Planck and BOSS in combination with other late-time probes.}
 \label{tab:qcmb}
\end{table}

As expected and apparent from the posteriors, we can see that Planck gives precise measurements on $\omega_b$, $\omega_{cdm}$, $\ln (10^{10}A_{s })$ and $n_s$, while constraints on $H_0$ or $\Omega_m$ are obtained by the combination with late-time probes, that break the degeneracy in the $H_0 - \Omega_m$ plane present in the CMB. 
As discussed in the previous subsection, $w$ is mainly measured thanks to low-redshift measurements.
However, the constraints on $w$ are better when adding Planck since the precise measurements of the other cosmological parameters by Planck helps to further break the degeneracies.

\subsection{$w$CDM with $w \geq -1$} 
From an effective field theory point of view, there is no known theory, at least to us, that can realize $w<-1$ with $c_s^2 \rightarrow 1$. 
As discussed in previous sections, such theory has a negative kinetic term. 
For a theory with no Lorentz-violating UV cutoff, the scalar perturbations are unstable, and the vacuum decays into gravitons at an infinite rate~\cite{Cline:2003gs}.
Therefore, $w<-1$ would either need some other, physical, motivation or one can posit that $w$ is not allowed to be smaller than $-1$ in $w$CDM.
By doing so, we get the results depicted in Fig.~\ref{fig:qcdm_vs_wcdm} obtained by fitting BOSS data on $w$CDM with a BBN prior and a flat prior $w \geq -1$. 
We see that the results differ substantially from the ones obtained without a prior on $w$. 
In particular, the degeneracy line $w-H_0$, open when allowing $w$ to vary below $-1$, can not be exploited to lift $H_0$ to higher values than the one found in $\Lambda$CDM analyzing CMB or LSS data.

\begin{figure}[h]
	\centering
	\includegraphics[width=0.99\textwidth]{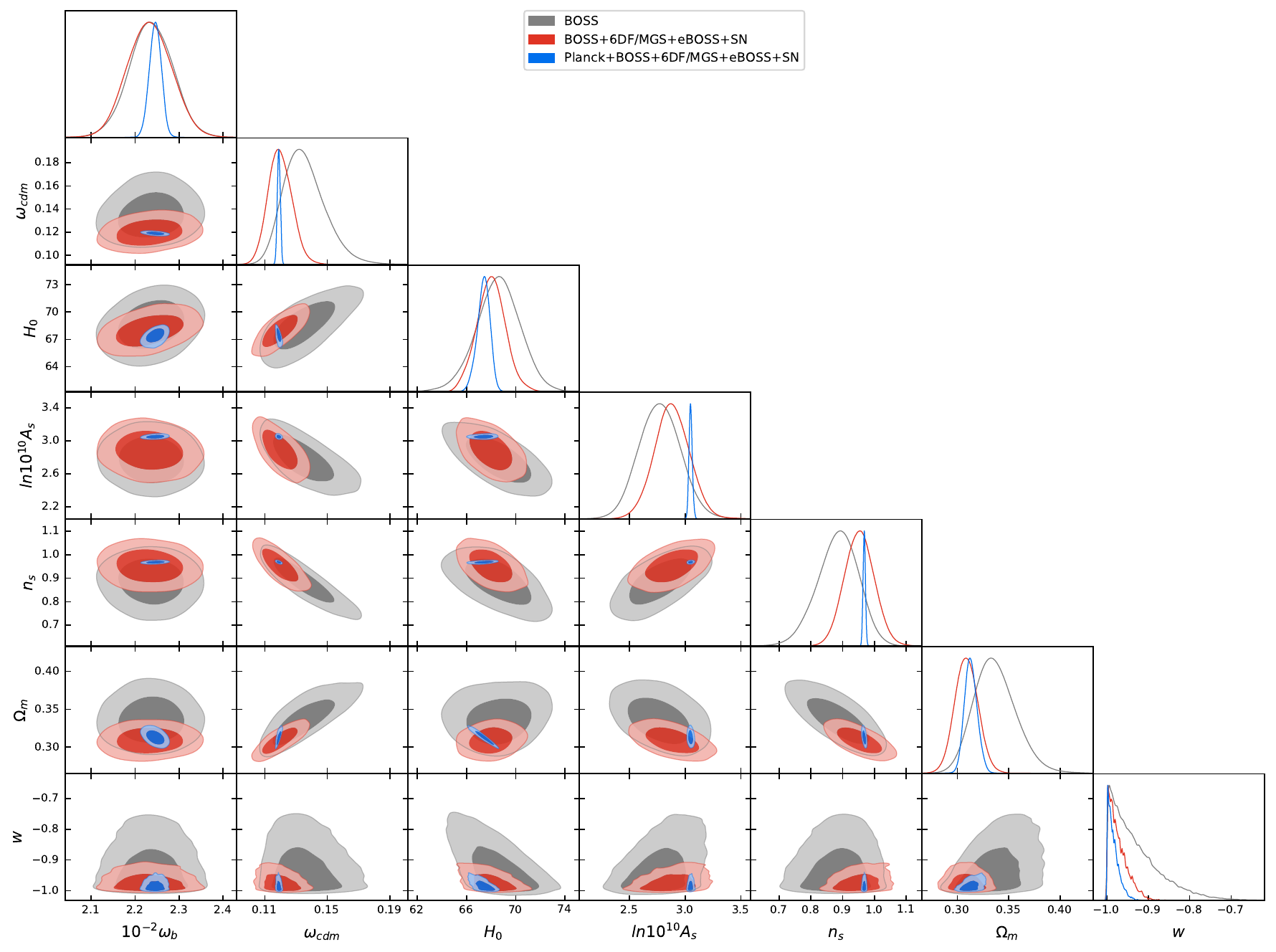}
	\caption{\small Triangle plots obtained by fitting smooth quintessence to BOSS in combination with other late-time probes, and to Planck, with a prior $w \geq -1$. When not fitting with Planck, we use a BBN prior. }
	\label{fig:wcdm}
\end{figure}

In Fig.~\ref{fig:wcdm} and Table~\ref{tab:wcdm}, we show the results obtained fitting BOSS, and in combination with BAO measurements from 6DF/MGS and eBOSS, with Pantheon SN, and with Planck data, on $w$CDM with a BBN prior and a prior $w \geq -1$.
{BOSS data alone gives a mild constraint $-1\leq w < -0.91$ at 68\% C.L. ($-1\leq w < - 0.81$ at $95\%$ C.L.).
Adding BAO information and Pantheon SN, the constraints on $H_0$ and especially $\Omega_m$ improve, giving the much stronger constraint $-1\leq w < - 0.96$ at $68\%$ C.L. ($-1\leq w < - 0.93$ at $95\%$ C.L.).
Finally, Planck improves this to $-1\leq w < - 0.979$ at $68\%$ C.L. ($-1\leq w < -0.956$ at $95\%$ C.L.), which means our Universe is consistent with a cosmological constant at $4\%$ precision.}

\begin{table}[h!]
\footnotesize
\centering
\begin{tabular}{|l|c|c||c|c||c|c|} 
 \hline 
		& \multicolumn{2}{c||}{BOSS} 					& \multicolumn{2}{c||}{\scriptsize BOSS+6DF/MGS+eBOSS+SN} 		& \multicolumn{2}{c|}{\scriptsize Planck+BOSS+6DF/MGS+eBOSS+SN} \\ \hline
 		 & best-fit & mean$\pm\sigma$ 					& best-fit & mean$\pm\sigma$ 						& best-fit & mean$\pm\sigma$ \\ \hline 
$100~\omega_{b }$ & $2.247$ & $2.236\pm 0.050$ 			&$2.275$ & $2.233\pm 0.050 $ 				&$2.247$ & $2.246\pm 0.013 $ \\
$\omega_{cdm }$ &$0.141$ & $0.135^{+0.010}_{-0.015} $ 	&$0.1211$ & $0.1198^{+0.0071}_{-0.0080}$			&$0.11849$ & $0.11896\pm 0.00094 $\\ 
$H_0$ &$70.25$ & $68.6\pm 1.8 $						&$68.45$ & $68.0\pm 1.2 $ 					&$67.98$ & $67.37^{+0.57}_{-0.45} $\\ 
$\ln (10^{10}A_{s })$ &$2.703$ & $2.77\pm 0.19 $ 	&$2.84$ & $2.88\pm 0.16 $					&$3.045$ & $3.050^{+0.013}_{-0.015} $\\ 
$n_{s }$ &$0.8754$ & $0.885^{+0.069}_{-0.058} $			&$0.95$ & $0.953\pm 0.047 $				&$0.97$ & $0.9681\pm 0.0037 $ \\
$\tau_{reio }$ & $-$ & $-$								&$-$ & $-$									&$0.0571$ & $0.0576^{+0.0067}_{-0.0079}$\\
$w$ & -0.9955 & $<-0.808 (2\sigma)$					&$-1.000$ & $<-0.927 (2\sigma)$	 				&$-0.998$ & $<-0.956 (2\sigma) $ \\ \hline
$\Omega_m$ &$0.3325$ & $0.337^{+0.017}_{-0.022} $ 		&$0.3084$ & $0.309\pm 0.011 $ 				&$0.3065$ & $0.3131^{+0.0056}_{-0.0066}$\\ 
$\sigma_8$ &$0.7345$ & $0.728\pm 0.047 $			&$0.733$ & $0.740^{+0.044}_{-0.050} $			&$0.8065$ & $0.8054\pm 0.0072 $\\ 
\hline 
 \end{tabular} \\
 \caption{Results obtained by fitting smooth quintessence to BOSS in combination with other late-time probes, and to Planck, with a prior $w \geq -1$. When not fitting with Planck, we use a BBN prior. 
For $w$, we quote the $95\%$ confidence bound instead of the $68\%$ confidence interval. }
 \label{tab:wcdm}
\end{table}
\pagebreak

\section*{Acknowledgments}
It is a pleasure to thank Diogo Braganca and Matthew Lewandowski for useful discussions.
YD is grateful for the kind hospitality at the Stanford Institute for Theoretical Physics (SITP)
at Stanford University and at the Kavli Institute for Particle Astrophysics and Cosmology (KIPAC) at SLAC National Accelerator Laboratory. LS is partially supported by Simons Foundation Origins of the Universe program (Modern Inflationary Cosmology collaboration) and LS by NSF award 1720397. 
PZ thanks Romane Schick for hospitality during completing of this work. 
Part of the analysis was performed on the Sherlock cluster at the Stanford University, for which we thank the support team, and part on the computer clusters LINDA \& JUDY in the particle cosmology group at the University of Science and Technology of China.

\begin{appendix}
\section{Green's functions}\label{appendixa}
At linear order, the time dependence is completely captured by the growth factor, defined as the solution of~\cite{Lewandowski:2014rca,Donath:2020abv}:
\be\label{delta-m34}
\frac{d^2}{d\ln a^2}\bigg(\frac{D}{H}\bigg)+\bigg(2+3\frac{d\ln H}{d\ln a}-\frac{d\ln C}{d\ln a}\bigg)\frac{d}{d\ln a}\bigg(\frac{D}{H}\bigg)=0 \ ,
\ee
The equation has two solutions, a growing mode
\be\label{D+-clus}
D_{+}(a)=\frac52 \int^a_0C(\ta)\Omega_{m}(\ta)\frac{H(a)}{H(\ta)}d\ta,
\ee
and a decaying mode
\be
D_{-}(a)=\frac{H(a)}{H_0\Omega_{m,0}^{1/2}} \ .
\ee
From these, we get the linear growth rates $f_{\pm}\equiv \frac{d\ln D_{\pm}}{d\ln a}$, given as
\be
f_{+} ( a ) =\bigg(\frac52\frac{a}{D_{+} ( a ) }-\frac32\bigg)\Omega_{m}(a)C(a) \, ,
\ee
and
\be
f_{-}(a)=-\frac32\Omega_{m}(a)C(a) \, ,
\ee
where 
\be
\Omega_{m}(a)\equiv \Omega_{m,0} \frac{H_0^2}{H(a)^2} a^{-3} \ , \hspace{.3in} \Omega_D(a) \equiv \Omega_{D,0} \frac{H_0^2}{H(a)^2} a^{-3 ( 1 + w) }
\ee
are the fractional matter and dark energy densities in terms of their present-day values $\Omega_{m,0}$ and $\Omega_{d,0}$.
To construct higher order solutions, it is useful to define Green's functions, coming from equations \eqref{eq:master1} and \eqref{eq:master2}:
\begin{align}
&a \frac{d G^{\delta}_{\sigma}(a,\ta)}{da}-f_{+}(a)G^{\theta}_{\sigma}(a,\ta)=\lambda_{\sigma}\delta(a-\ta), \label{Green} \\
&a \frac{d G^{\theta}_{\sigma}(a,\ta)}{da}-f_{+}(a)G^{\theta}_{\sigma}(a,\ta)-\frac{f_{-}(a)}{f_{+}(a)}\bigg(G^{\theta}_{\sigma}(a,\ta)-G^{\delta}_{\sigma}(a,\ta)\bigg)=(1-\lambda_{\sigma})\delta(a-\ta),
\end{align}
where $\sigma \in \{1,2\}$, $\lambda_1 = 1$ and $\lambda_2 = 0$.
Explicitly they are given by
\begin{align}
&G^{\delta}_1(a,\ta)=\frac{1}{\ta W(\ta)}\bigg(\frac{d D_{-}(\ta)}{d\ta}D_{+}(a)-\frac{d D_{+}(\ta)}{d\ta}D_{-}(a)\bigg) {\Theta}(a-\ta) \label{gdelta} \ ,\\
&G^{\delta}_2(a,\ta)=\frac{f_{+}(\ta)/\ta^2}{W(\ta)}\bigg(D_{+}(\ta)D_{-}(a)-D_{-}(\ta)D_{+}(a)\bigg){\Theta}(a-\ta) \ , \\
&G^{\theta}_1(a,\ta)=\frac{a/\ta}{f_{+}(a)W(\ta)}\bigg(\frac{d D_{-}(\ta)}{d\ta}\frac{d D_{+}(a)}{d a}-\frac{d D_{+}(\ta)}{d\ta}\frac{d D_{-}(a)}{d a}\bigg) {\Theta}(a-\ta) \ ,\\
&G^{\theta}_2(a,\ta)=\frac{f_{+}(\ta)a/\ta^2}{f_{+}(a)W(\ta)}\bigg(D_{+}(\ta)\frac{d D_{-}(a)}{d a}-D_{-}(\ta)\frac{d D_{+}(a)}{d a}\bigg) {\Theta}(a-\ta) \ , \label{gtheta}
\end{align}
where $\Theta (a-\tilde a)$ is the Heaviside step function, $W(\ta)$ is the Wronskian of $D_+$ and $D_-$: 
\be
W(\ta)=\frac{dD_{-}(\ta)}{d\ta}D_{+}(\ta)-\frac{d D_{+}(\ta)}{d\ta}D_{-}(\ta) \ ,
\ee
and we impose the boundary conditions \begin{align} \label{bound}
& G^{\delta}_\sigma(a,\tilde a) = 0 \quad \quad \text{and} \quad\quad G^{\theta}_\sigma(a, \tilde a)=0 \quad \quad \text{for} \quad \quad \tilde a > a \ , \\
&G^\delta_\sigma ( \tilde a , \tilde a ) = \frac{\lambda_\sigma}{\tilde a} \quad \hspace{.06in} \text{and} \hspace{.2in} \quad G^{\theta}_{\sigma} ( \tilde a , \tilde a ) = \frac{(1 - \lambda_\sigma)}{\tilde a} . \label{bound2}
\end{align}
At second order, the resulting time-dependent functions are given by
\begin{align}
\label{2nd-solution} 
\mG^{\delta}_{\sigma}(a)&=\int^{1}_0 G^{\delta}_{\sigma}(a,\ta)\frac{f_{+}(\ta) D_{+}^2(\ta) }{ C(\ta)D_+^2 (a) }d\ta \ , \\
\mG^{\theta}_{\sigma}(a)&=\int^{ 1}_0G^{\theta}_{\sigma}(a,\ta) \frac{f_{+}(\ta) D_{+}^2(\ta) }{C(\ta) D_+^2 (a ) }d\ta, \label{2nd-solution2} 
\end{align}
for $\sigma = 1,2$.
At third order we have
\begin{align}\label{3rd-solution} 
\mU^{\delta}_{\sigma}(a)=\int^{{1}}_0 G^{\delta}_{1}(a,\ta)\frac{f_{+}(\ta)D^3_{+}(\ta)}{ C(\ta)D^3_+ (a) }\mG^{\delta}_{\sigma}(\ta)d\ta,\\
\mU^{\theta}_{\sigma}(a)=\int^{{1}}_0 G^{\theta}_{1}(a,\ta)\frac{f_{+}(\ta)D^3_{+}(\ta)}{C(\ta) D^3_+ (a) }\mG^{\delta}_{\sigma}(\ta)d\ta,\\
\mV^{\delta}_{\sigma\tilde\sigma}(a)=\int^{{ 1}}_0 G^{\delta}_{\tilde\sigma}(a,\ta)\frac{f_{+}(\ta)D^3_{+}(\ta)}{C(\ta) D^3_+ (a) }\mG^{\theta}_{\sigma}(\ta)d\ta,\\
\mV^{\theta}_{\sigma\tilde\sigma}(a)=\int^{{1}}_0 G^{\theta}_{\tilde\sigma}(a,\ta)\frac{f_{+}(\ta)D^3_{+}(\ta)}{C(\ta) D^3_+ (a) }\mG^{\theta}_{\sigma}(\ta)d\ta.
\label{3rd-solutionlast}
\end{align}
The degeneracies pointed out in~\eqref{eq:degen} result from the following identities:
\bea \label{eq:timeidents}
\mG_1^\delta+\mG_2^\delta=\mG_1^\theta+\mG_2^\theta&=&\mG\\ \nonumber
\mV_{11}^\delta+\mV_{21}^\delta &=& \mU_1^\delta+ \mU_2^\delta\\ \nonumber
\mV_{11}^\theta+\mV_{21}^\theta &=& \mU_1^\theta+ \mU_2^\theta\\ \nonumber
\mV_{\sigma1}^\delta+\mV_{\sigma2}^\delta &=&\mV_{\sigma1}^\theta+\mV_{\sigma2}^\theta \\ \nonumber
\mV_{11}^\delta+\mV_{21}^\delta+\mV_{12}^\delta+\mV_{22}^\delta&=& \fr{\mG^2}{2}\\ \nonumber
\mV_{11}^\theta+\mV_{21}^\theta+\mV_{12}^\theta+\mV_{22}^\theta &=&\fr{\mG^2}{2}\\ \nonumber
\mU_1^{\delta}-\mV_{22}^{\delta}&=&\fr{\mG}{2} \left(\mG_1^{\delta}-\mG_2^{\delta}\right)		\\ \nonumber
\mU_1^{\theta}-\mV_{22}^{\theta}&=&\fr{\mG}{2} \left(\mG_1^{\theta}-\mG_2^{\theta}\right)			
\eea
where again $\sigma\in \{1,2\}$. One can derive these relations using \eqref{2nd-solution}-\eqref{3rd-solutionlast} and the fact that
\bea G^{\delta}_1(a,\ta)+G^{\delta}_2(a,\ta)&=&G^{\theta}_1(a,\ta)+G^{\theta}_2(a,\ta) = \fr{D_{+}(a)}{\ta D_{+}(\ta)}\Theta(a-\ta)\\
G^{\delta}_1(a,\ta)-G^{\theta}_1(a,\ta)&=& \frac{W(a)}{\ta W(\ta)}\fr{D_+'(\ta)}{D_+'(a)} {\Theta}(a-\ta) . 
\eea
Furthermore, for the derivation of some of the flow terms in Appendix \ref{appendixc} it is important to use the following relations:
\begin{gather}
	\label{tintegral}
	\mV^{\delta}_{\sigma1}(a)+\mV^{\delta}_{\sigma2}(a)=\int^{{ a}}_0 \frac{D'_{+}(\ta)D_{+}(\ta)}{C(\ta)D^2_+ (a) }\mG^{\theta}_{\sigma}(\ta)d\ta \, , \\
	\label{Gint}
	\mG = \int^a_0 \fr{D_+'(\ta)}{C(\ta)D_{+}(a)}d\ta \, , \\
	\int^a_0 \mG(\ta)\fr{D_+'(\ta)D_+(\ta)}{C(\ta)D_{+}(a)^2}d\ta=\fr{\mG^2}{2} \, .
\end{gather}

\section{Bias operators, Halo kernels and degeneracy of halo bias parameters}\label{appendixb}

In this section we quickly wish to outline how we get from equation \eqref{eq:euler_bias_raw} to \eqref{eq:delta_h_CoI_t}.
First, we define the operators that appear in \eqref{eq:euler_bias_raw}. In the exact same way as in \cite{Donath:2020abv}, we
follow the approach used by~\cite{McDonald:2009dh}, generalized to exact time dependence. Using
\be\label{eta}
\eta(\vec x,t)=\theta(\vec x,t)- \delta(\vec x,t)\, ,
\ee
we define
\beq 
s_{ij}(\vec{x},a)=\mathcal{D}_{ij}\delta(\vec{x},a) \quad \textmd{and} \quad t_{ij}(\vec{x},a)=\mathcal{D}_{ij}\eta(\vec{x},a),
\eeq
where $\mathcal{D}_{ij} = \fr{\partial_i \partial_j}{\partial^2}-\fr{1}{3}\delta_{ij}$.
Then we get the contractions
\bea\label{eq:operators}
&&s^2(\xfl,a)= s_{ij}(\xfl,a) s^{ij}(\xfl,a)\ , \quad s^3(\xfl,a)=s_{ij}(\xfl,a) s^{il}(\xfl,a)s_l{}^{j}(\xfl,a)\ ,\\ \nonumber
&& \epsilon s(\xfl,a)=\epsilon_{ij}(\xfl,a)s^{ij}(\xfl,a), \quad \epsilon t(\xfl,a)=\epsilon_{ij}(\xfl,a)t^{ij}(\xfl,a)\ , \\ \nonumber
&& st(\xfl,a)=s_{ij}(\xfl,a) t^{ij}(\xfl,a)\, .
\eea
Furthermore, $\psi$ is given by
\bea 
\psi(\vec{x},a) = \theta(\vec{x},a)-\delta(\vec{x},a)-\left(\mG^{\delta}_{1}(a)-\mG^{\theta}_{1}(a)\right)\left(s^2(\vec{x},a)-\frac{2}{3}\delta^2(\vec{x},a)\right),
\eea	
so that it only starts at third order.
The $\epsilon$ and $\epsilon_{ij}$ are stochastic operators, which are uncorrelated with the density field.
Their correlation functions will not depend on the initial power spectrum and contain all terms allowed by rotational invariance in a derivative expansion, that is $\langle \epsilon(\vec{k}) \epsilon(\vec{k}') \rangle' =  c_0 + c_1 \frac{k^2}{k_{\rm M}^2} + \dots $, $\langle \epsilon(\vec{k}) \epsilon_{ij}(\vec{k}') \rangle' = d_0 \delta_{ij} + d_1 \frac{k^2}{k_{\rm M}^2} \delta_{ij} + d_2 \frac{k^i k^j}{k_{\rm M}^2} + \dots $. The $\langle \dots \rangle'$ notation means that the correlation is stripped of the momentum-conserving Dirac delta.

One can show that all operators in \eqref{eq:euler_bias_raw}, including the flow terms, {up to cubic order in the fluctuations}, can be expressed as linear combinations of the following nine momentum kernels (see~\cite{Donath:2020abv},~we can call the `exact-time' basis or the `Greek' basis):
\begin{align}\label{alphakernels}
&\mathbb{I} = 1\\
&\alpha(\vq_1,\vq_2) = 1+\fr{\vq_1\cdot\vq_2}{q_1^2}\\
&\beta(\vq_1,\vq_2) = \fr{(\vq_1+\vq_2)^2\vq_1\cdot\vq_2}{2q_1^2q_2^2}\\
&\alpha^1(\vq_1,\vq_2,\vq_3)=\alpha(\vq_3,\vq_1+\vq_2)\alpha_s(\vq_1,\vq_2), \\ &\alpha^2(\vq_1,\vq_2,\vq_3)=\alpha(\vq_3,\vq_1+\vq_2)\beta(\vq_1,\vq_2), \\
&\beta^1(\vq_1,\vq_2,\vq_3)=2\beta(\vq_3,\vq_1+\vq_2)\alpha_s(\vq_1,\vq_2), \\ &\beta^2(\vq_1,\vq_2,\vq_3)=2\beta(\vq_3,\vq_1+\vq_2)\beta(\vq_1,\vq_2), \\
&\gamma^1(\vq_1,\vq_2,\vq_3)=\alpha(\vq_1+\vq_2,\vq_3)\alpha_s(\vq_1,\vq_2), \\ &\gamma^2(\vq_1,\vq_2,\vq_3)=\alpha(\vq_1+\vq_2,\vq_3)\beta(\vq_1,\vq_2).
\end{align}

The resulting redefinitions of parameters that appear in \eqref{eq:euler_bias_k} and \eqref{eq:delta_h_CoI_t} are given by
\begin{align}\label{third order final coefficients}
&c_{\alpha,(2)}=\mG\cdot c_{\delta,1}-c_{\delta_2,\mG_2^\delta}-c_{s^2,1}\\ \nonumber
&c_{\beta,(2)}=c_{\delta_2,\mG_2^\delta}+c_{s^2,1}\\ \nonumber
&c_{\mathbb{I},(2)} =-\mG\cdot c_{\delta,1}+ c_{\delta_2,\mG_1^\delta}+c_{\delta_2,\mG_2^\delta}+c_{\delta^2,1}+\sfrac{2}{3}c_{s^2,1}\\ \nonumber
&c_{\alpha_1,(3)} =\sfrac{1}{2}\mG^2\cdot c_{\delta,1}-\mG \cdot c_{\delta_2,\mG_2^\delta}-\sfrac{1}{2}\left(c_{\delta,\mG^{\delta}_{1}}-c_{\delta,\mG^{\delta}_{2}}\right)+c_{\delta,\mU^{\delta}_{1}}-\mG \cdot c_{s^2,{1}}+c_{s^2,\mG^{\delta}_{2}}-\sfrac{1}{2}\left(c_{st,\mG^{\theta}_{1}}-c_{st,\mG^{\delta}_{1}}\right)\\ \nonumber
&\qquad\qquad+c_{\psi,\mU^{\theta}_{1}}-c_{\psi,\mU^{\delta}_{1}}+ c_{\psi,\mG^{\delta}_{1}}+\sfrac{1}{2}c_{s^3}\\ \nonumber
&c_{\alpha_2,(3)} =\mG \cdot c_{\delta_2,\mG^{\delta}_{2}}-c_{\delta,\mG^{\delta}_{2}}+c_{\delta,\mU^{\delta}_{2}}+\mG \cdot c_{s^2,{1}}-c_{s^2,\mG^{\delta}_{1}}-2\,c_{s^2,\mG^{\delta}_{2}}-\sfrac{1}{2}\left(c_{st,\mG^{\theta}_{2}}-c_{st,\mG^{\delta}_{2}}\right)\\ \nonumber
&\qquad\qquad+c_{\psi,\mU^{\theta}_{2}}-c_{\psi,\mU^{\delta}_{2}}+ c_{\psi,\mG^{\delta}_{2}}-c_{s^3}\\ \nonumber
&c_{\beta_1,(3)} =c_{\delta,\mV^{\delta}_{12}}+c_{s^2,\mG^{\delta}_{1}}+\sfrac{1}{2}\left(c_{st,\mG^{\theta}_{1}}-c_{st,\mG^{\delta}_{1}}\right)+c_{\psi,\mV^{\theta}_{12}}-c_{\psi,\mV^{\delta}_{12}}- c_{\psi,\mG^{\delta}_{1}}\\ \nonumber
&c_{\beta_2,(3)} =c_{\delta,\mV^{\delta}_{22}}+c_{s^2,\mG^{\delta}_{2}}+\sfrac{1}{2}\left(c_{st,\mG^{\theta}_{2}}-c_{st,\mG^{\delta}_{2}}\right)+c_{\psi,\mV^{\theta}_{22}}-c_{\psi,\mV^{\delta}_{22}}- c_{\psi,\mG^{\delta}_{2}}+\sfrac{1}{2}c_{s^3}\\ \nonumber
&c_{\gamma_1,(3)} =\left(\mV_{11}^{\delta}+\mV_{12}^{\delta}\right)c_{\delta,1}-c_{\delta,\mV^{\delta}_{12}}-c_{s^2,\mG^{\delta}_{1}}-\sfrac{1}{2}\left(c_{st,\mG^{\theta}_{1}}-c_{st,\mG^{\delta}_{1}}\right)+c_{\psi,\mV^{\theta}_{11}}-c_{\psi,\mV^{\delta}_{11}}+ c_{\psi,\mG^{\delta}_{1}}\\ \nonumber
&c_{\gamma_2,(3)} =\left(\mV_{21}^{\delta}+\mV_{22}^{\delta}\right)c_{\delta,1}-c_{\delta,\mV^{\delta}_{22}}-c_{s^2,\mG^{\delta}_{2}}-\sfrac{1}{2}\left(c_{st,\mG^{\theta}_{2}}-c_{st,\mG^{\delta}_{2}}\right)+c_{\psi,\mV^{\theta}_{21}}-c_{\psi,\mV^{\delta}_{21}}+ c_{\psi,\mG^{\delta}_{2}}-\sfrac{1}{2}c_{s^3}\\ \nonumber
&c_{\alpha,(3)} =-\sfrac{3}{2}\mG^2\cdot c_{\delta,1}-\left(\mV_{11}^{\delta}+\mV_{12}^{\delta}\right)c_{\delta,1}+c_{\delta,\mV_{11}^{\delta}}+c_{\delta,\mV_{12}^{\delta}} +\mG \cdot\left(2\,c_{\delta_2,\mG_1^{\delta}}+3\,c_{\delta_2,\mG_2^{\delta}}\right)-\sfrac{1}{2}\left(c_{\delta,\mG_1^{\delta}}+3\,c_{\delta,\mG_2^{\delta}}\right)\\ \nonumber
&\qquad\qquad+2\,\mG \cdot c_{\delta^2,1}-2\,c_{\delta^2,\mG^{\delta}_{2}}+\sfrac{7}{3}\mG \cdot c_{s^2,1}-c_{s^2,\mG^{\delta}_{1}}-\sfrac{7}{3}c_{s^2,\mG^{\delta}_{2}}+\sfrac{2}{3}\left(c_{st,\mG^{\theta}_{1}}-c_{st,\mG^{\delta}_{1}}\right)-c_{\delta s^2}-\sfrac{1}{2}c_{s^3}\\ \nonumber
&c_{\beta,(3)} =-\left(\mV_{21}^{\delta}+\mV_{22}^{\delta}\right)c_{\delta,1}+c_{\delta,\mV_{21}^{\delta}}+c_{\delta,\mV_{22}^{\delta}} -\mG \cdot c_{\delta_2,\mG_2^{\delta}}+c_{\delta,\mG_2^{\delta}}\\ \nonumber
&\qquad\qquad+2\,c_{\delta^2,\mG^{\delta}_{2}}-\mG \cdot c_{s^2,1}+c_{s^2,\mG^{\delta}_{1}}+\sfrac{7}{3}c_{s^2,\mG^{\delta}_{2}}+\sfrac{2}{3}\left(c_{st,\mG^{\theta}_{2}}-c_{st,\mG^{\delta}_{2}}\right)+c_{\delta s^2}+\sfrac{1}{2}c_{s^3}\\ \nonumber
&c_{\mathbb{I},(3)} = \mG^2 \cdot c_{\delta,1}-2\,\mG\left(c_{\delta_2,\mG_1^{\delta}}+c_{\delta_2,\mG_2^{\delta}}\right)+c_{\delta,\mG_1^{\delta}}+c_{\delta,\mG_2^{\delta}}-2\,\mG \cdot c_{\delta^2,1}+2\left(c_{\delta^2,\mG^{\delta}_{1}}+c_{\delta^2,\mG^{\delta}_{2}}\right)\\ \nonumber
&\qquad\qquad-\sfrac{4}{3}\mG \cdot c_{s^2,1}+\sfrac{4}{3}\left(c_{s^2,\mG^{\delta}_{1}}+c_{s^2,\mG^{\delta}_{2}}\right)+\sfrac{2}{9}c_{s^3}+\sfrac{2}{3}c_{\delta s^2}+c_{\delta^3}
\end{align} 
where the coefficients that appear here are the symbolic integrals over the time-dependent functions defined in Appendix~\ref{appendixa} that come from the expansion~\eqref{eq:euler_bias_raw}. \\
They read 
\begin{alignat}{3}
\label{coefficientraw}
&c_{\delta,1}(a) &&= \int^a \frac{da'}{a'}c_{\delta}(a,a')\frac{D_{+}(a')}{D_{+}(a)},\\ \nonumber
&c_{\delta_2,\mG^{\delta}_{\sigma}}(a) &&= \int^a \frac{da'}{a'}c_\delta(a,a')\frac{D_{+}(a')^2}{D_{+}(a)^2}\mG^{\delta}_{\sigma}(a') 
&&\\ \nonumber
&c_{s^2,1}(a) &&= \int^a \frac{da'}{a'}c_{s^2}(a,a') \frac{D_{+}(a')^2}{D_{+}(a)^2}
&&c_{\delta^2,1}(a) = \int^a \frac{da'}{a'}c_{\delta^2}(a,a') \frac{D_{+}(a')^2}{D_{+}(a)^2},\\ \nonumber
&c_{s^3}(a) &&= \int^a \frac{da'}{a'}c_{s^3}(a,a')\frac{D_{+}(a')^3}{D_{+}(a)^3}
&&c_{\delta^3}(a) = \int^a \frac{da'}{a'}c_{\delta^3}(a,a')\frac{D_{+}(a')^3}{D_{+}(a)^3}\\ \nonumber
&c_{\delta,\mU^{\delta}_{\sigma}}(a) &&= \int^a \frac{da'}{a'}c_\delta(a,a')\frac{D_{+}(a')^3}{D_{+}(a)^3}\mU^{\delta}_{\sigma}(a') 
&&c_{\delta,\mG^{\delta}_{\sigma}}(a) = \int^a \frac{da'}{a'}c_\delta(a,a')\frac{D_{+}(a')^3}{D_{+}(a)^3}\mG(a')\mG^{\delta}_{\sigma}(a')\\ \nonumber
&c_{\delta,\mV^{\delta}_{\sigma \tilde{\sigma}}}(a) &&= \int^a \frac{da'}{a'}c_\delta(a,a')\frac{D_{+}(a')^3}{D_{+}(a)^3}\mV^{\delta}_{\sigma \tilde{\sigma}}(a')
&&c_{\delta s^2}(a) = \int^a \frac{da'}{a'}c_{\delta s^2}(a,a')\frac{D_{+}(a')^3}{D_{+}(a)^3}\\ \nonumber
&c_{\delta^2,\mG^{\delta}_{\sigma}}(a) &&= \int^a \frac{da'}{a'}c_{\delta^2}(a,a')\frac{D_{+}(a')^3}{D_{+}(a)^3}\mG^{\delta}_{\sigma}(a') 
&&c_{s^2,\mG^{\delta}_{\sigma}}(a) = \int^a \frac{da'}{a'}c_{s^2}(a,a')\frac{D_{+}(a')^3}{D_{+}(a)^3}\mG^{\delta}_{\sigma}(a') \\ \nonumber
&c_{st,\mG^{\delta}_{\sigma}}(a) &&= \int^a \frac{da'}{a'}c_{st}(a,a')\frac{D_{+}(a')^3}{D_{+}(a)^3}\mG^{\delta}_{\sigma}(a') &&c_{st,\mG^{\theta}_{\sigma}}(a) = \int^a \frac{da'}{a'}c_{st}(a,a')\frac{D_{+}(a')^3}{D_{+}(a)^3}\mG^{\theta}_{\sigma}(a')\\ \nonumber
&c_{\psi,\mU^{\delta}_{\sigma}}(a) &&= \int^a \frac{da'}{a'}c_\psi(a,a')\frac{D_{+}(a')^3}{D_{+}(a)^3}\mU^{\delta}_{\sigma}(a') 
&&c_{\psi,\mV^{\delta}_{\sigma \tilde{\sigma}}}(a) = \int^a \frac{da'}{a'}c_\psi(a,a')\frac{D_{+}(a')^3}{D_{+}(a)^3}\mV^{\delta}_{\sigma \tilde{\sigma}}(a')\\ \nonumber
&c_{\psi,\mU^{\theta}_{\sigma}}(a) &&= \int^a \frac{da'}{a'}c_\psi(a,a')\frac{D_{+}(a')^3}{D_{+}(a)^3}\mU^{\theta}_{\sigma}(a') 
&&c_{\psi,\mV^{\theta}_{\sigma \tilde{\sigma}}}(a) = \int^a \frac{da'}{a'}c_\psi(a,a')\frac{D_{+}(a')^3}{D_{+}(a)^3}\mV^{\theta}_{\sigma \tilde{\sigma}}(a')\\ \nonumber
&c_{\psi\mG^{\delta}_{\sigma}}(a) &&= \int^a \frac{da'}{a'}c_\psi(a,a')\frac{D_{+}(a')^3}{D_{+}(a)^3}\mG^{\delta}_{\sigma}(a')\big(\mG^{\delta}_{1}(a')&&-\mG^{\theta}_{1}(a')\big)\\ \nonumber
\end{alignat}

For completeness, we here explicitly write the $\mathbb{C}_i$ operators that appear in \eqref{eq:delta_h_CoI_t}:
\bea \label{eq:Ci_ops}
{}^*\mathbb{C}^{(1)}_{\delta} (\vq_1)&=& 1\\ \nonumber
{}^*\mathbb{C}^{(2)}_{\delta} (\vq_1,\vq_2)&=& \beta(\vq_1,\vq_2)\\ \nonumber
{}^*\mathbb{C}^{(2)}_{\alpha}(\vq_1,\vq_2)&=& \alpha(\vq_1,\vq_2)-\beta(\vq_1,\vq_2)\\ \nonumber 
{}^*\mathbb{C}^{(2)}_{\mathbb{I}} (\vq_1,\vq_2)&=& 1\\ \nonumber
{}^*\mathbb{C}^{(3)}_{\delta}(\vq_1,\vq_2,\vq_3)&=& -\fr{3}{14}\alpha_1(\vq_1,\vq_2,\vq_3)+\fr{3}{7}\alpha_2(\vq_1,\vq_2,\vq_3) +\fr{2}{7}\beta_2(\vq_1,\vq_2,\vq_3) + \fr{3}{14}\gamma_1(\vq_1,\vq_2,\vq_3)\\ \nonumber
{}^*\mathbb{C}^{(3)}_{\alpha_1}(\vq_1,\vq_2,\vq_3)&=&\alpha_1(\vq_1,\vq_2,\vq_3)-\alpha_2(\vq_1,\vq_2,\vq_3)\\ \nonumber
{}^*\mathbb{C}^{(3)}_{\beta_1}(\vq_1,\vq_2,\vq_3)&=&-\alpha_2(\vq_1,\vq_2,\vq_3) + \beta_1(\vq_1,\vq_2,\vq_3) - \gamma_1(\vq_1,\vq_2,\vq_3)\\ \nonumber
{}^*\mathbb{C}^{(3)}_{\gamma_2}(\vq_1,\vq_2,\vq_3)&=&-\alpha_1(\vq_1,\vq_2,\vq_3)+2\alpha_2(\vq_1,\vq_2,\vq_3) - \beta_2(\vq_1,\vq_2,\vq_3) + \gamma_2(\vq_1,\vq_2,\vq_3)\\ \nonumber
{}^*\mathbb{C}^{(3)}_{\alpha}(\vq_1,\vq_2,\vq_3)&=&\alpha(\vq_1,\vq_2)-\beta(\vq_1,\vq_2)\\ \nonumber
{}^*\mathbb{C}^{(3)}_{\beta}(\vq_1,\vq_2,\vq_3)&=&\beta(\vq_1,\vq_2)\\ \nonumber
{}^*\mathbb{C}^{(3)}_{\mathbb{I}}(\vq_1,\vq_2,\vq_3)&=&1\\ \nonumber
{}^*\mathbb{C}^{(3)}_{Y}(\vq_1,\vq_2,\vq_3)&=&-\alpha_1(\vq_1,\vq_2,\vq_3)+2\alpha_2(\vq_1,\vq_2,\vq_3) - \beta_2(\vq_1,\vq_2,\vq_3) + \gamma_1(\vq_1,\vq_2,\vq_3),\\ \nonumber
\eea
where the $\mathbb{C}_i$ are related to the ${}^*\mathbb{C}_i$ by
\be\label{kernelnotation}
\mathbb{C}_i^{(n)}(\vk,a) = \int\frac{d^3q_1}{(2\pi)^{3}}\ldots\frac{d^3q_n}{(2\pi)^{3}}(2\pi)^{3}\delta_{D}(\vk-\vq_1-\ldots-\vq_n) \ {}^*\mathbb{C}_i^{(n)}(\vq_1,...,\vq_n)\ \delta^{(1)}_{\vq_1}(a)\ldots\delta^{(1)}_{\vq_n}(a).
\ee

\section{Deriving flow terms}\label{appendixc}
We here derive the flow terms coming from the Taylor expansion 
\bea\label{fluidtaylor}
&&\delta(\xfl(a,a'),a')=\delta(\vec x,a')-\d_i\delta(x,a')\int_{a'}^a 	\fr{da''}{a''^2H(a'')} \; v^i(\vec x,a'')\\ \nonumber
&&\quad \quad\quad\quad\quad\quad+\frac{1}{2}\d_i\d_j \delta(x,a')\int_{a'}^a 	\fr{da''}{a''^2H(a'')}\; v^i(\vec x,a'')\int_{a'}^a 	\fr{da'''}{a'''^2H(a''')}\;v^j(\vec x,a''')\\ \nonumber
&&\quad \quad\quad \quad\quad \quad+\d_i\delta(x,a')\int_{a'}^a 	\fr{da''}{a''^2H(a'')}\; \d_j v^i(\vec x,a'') \int^a_{a''} 	\fr{da'''}{a'''^2H(a''')}\;v^j(\vec x,a''')+\ldots\ .
\eea 
In the bias expansion from \cite{Senatore:2014eva,Donath:2020abv} we integrate over time integral kernels such as $c_\delta(a,a')$, which we will be including in the following.
We will often us the former definition $v^i =-a^2 H\fr{\,D_+'}{D_+\,C}\fr{\d_i}{\d^2}\theta $, as well as the star notation from \eqref{kernelnotation}.

First, we expand the overdensity and velocity divergence perturbatively. Apart from $\delta^{(2)}$, the only second-order term is in the first line, which is given by
\bea\label{appendix-flow}
&& -\int^a \fr{da'}{a'}\; c_{\delta}(a,a')\; \d_i \delta^{(1)}(a') \int^a_{a'} \fr{da''}{a''^2H(a'')} v^{(1)}{}^i(a'')=\\ \nonumber
&&\quad = \int^a \fr{da'}{a'}\; c_{\delta}(a,a')\; \frac{D_{+}(a')}{D_{+}(a)}\d_i \delta^{(1)}(a) \int^a_{a'} da''\;\frac{D'_{+}(a'')}{C(a'')D_{+}(a)}\frac{\d^i}{\d^2}\theta^{(1)} (a)\\ \nonumber
&&\quad=\int^a \fr{da'}{a'}\; c_{\delta}(a,a')\; \frac{D_{+}(a')}{D_{+}(a)}\d_i \delta^{(1)}(a)\frac{\d^i}{\d^2}\theta^{(1)} (a)\left[\mG(a)-\frac{D_{+}(a')}{D_{+}(a)}\mG(a')\right]\\ \nonumber
&&\quad=\left[c_{\delta,1}(a)\mG(a)-c_{\delta_2,\mG^{\delta}_{1}}(a)-c_{\delta_2,\mG^{\delta}_{2}}(a)\right]\d_i \delta^{(1)}(a)\frac{\d^i}{\d^2}\theta^{(1)} (a)\\ \nonumber
&&\quad\stackrel{w\text{CDM}}{=}c_{\delta,12}(a)\d_i \delta^{(1)}(a) \frac{\d^i}{\d^2}\theta^{(1)} (a)\ .
\eea
Next, we take this same term with $\delta$ at second order and $v$ at first order. This gives
\bea
&& -\int^a \fr{da'}{a'}\; c_{\delta}(a,a')\; \d_i \delta^{(2)}(a') \int^a_{a'} \fr{da''}{a''^2H(a'')}\; v^{(1)}{}^i(a'')=\\ \nonumber
&&\quad=\int^a \fr{da'}{a'}\; c_{\delta}(a,a')\; \left[\mG(a)-\frac{D_{+}(a')}{D_{+}(a)}\mG(a')\right]\d_i \delta^{(2)}(a') \frac{\d^i}{\d^2}\theta^{(1)} (a)\ .
\eea
In Fourier space this reads
\bea
&&=\left[\mG(a)c_{\delta_2,\mG_1^\delta}-c_{\delta,\mG_1^\delta}\right]\left(\alpha^1(\vq_1,\vq_2,\vq_3)-\alpha(\vq_1,\vq_2)\right)\\ \nonumber
&&\quad+\left[\mG(a)c_{\delta_2,\mG_2^\delta}-c_{\delta,\mG_2^\delta}\right]\left(\alpha^2(\vq_1,\vq_2,\vq_3)-\beta(\vq_1,\vq_2)\right)\\ \nonumber
&&\quad\stackrel{\text{EdS}}{=}\left[c_{\delta,2}(a)-c_{\delta,3}(a)\right]{}^*[\d_i \delta^{(2)} \frac{\d^i}{\d^2}\theta^{(1)}]_{\vk} (a),
\eea
Again, from the same term, we can take $\delta$ at linear and $v$ at second order. We have
\bea
&& -\int^a \fr{da'}{a'}\; c_{\delta}(a,a')\; \d_i \delta^{(1)}(a') \int^a_{a'}\fr{da''}{a''^2H(a'')}\; v^{(2)}{}^i(a'')=\\ \nonumber
&&\quad=\int^a \fr{da'}{a'}\; c_{\delta}(a,a')\; \frac{D_{+}(a')}{D_{+}(a)}\d_i \delta^{(1)}(a) \int^a_{a'} da''\;\frac{D'_{+}(a'')}{C(a'')D_{+}(a'')}\frac{\d^i}{\d^2}\theta^{(2)} (a'')\ .
\eea
In terms of Fourier space kernels this reads
\bea
&&=\int^a \fr{da'}{a'}\; c_{\delta}(a,a')\; \frac{D_{+}(a')}{D_{+}(a)} \int^a_{a'} da''\;\frac{D'_{+}(a'')D_{+}(a'')}{C(a'')D_{+}(a)^2}\mG_1^\theta(a'')\left(\gamma^1(\vq_1,\vq_2,\vq_3)-\alpha(\vq_1,\vq_2)\right)\\ \nonumber
&&\quad+\int^a \fr{da'}{a'}\; c_{\delta}(a,a')\; \frac{D_{+}(a')}{D_{+}(a)} \int^a_{a'} da''\;\frac{D'_{+}(a'')D_{+}(a'')}{C(a'')D_{+}(a)^2}\mG_2^\theta(a'')\left(\gamma^2(\vq_1,\vq_2,\vq_3)-\beta(\vq_1,\vq_2)\right)\\ \nonumber
&&=\left[\left(\mV_{11}^\delta(a)+\mV_{12}^\delta(a)\right)c_{\delta,1}-c_{\delta,\mV_{11}^\delta}-c_{\delta,\mV_{12}^\delta}\right]\left(\gamma^1(\vq_1,\vq_2,\vq_3)-\alpha(\vq_1,\vq_2)\right)\\ \nonumber
&&\quad+\left[\left(\mV_{21}^\delta(a)+\mV_{22}^\delta(a)\right)c_{\delta,1}-c_{\delta,\mV_{21}^\delta}-c_{\delta,\mV_{22}^\delta}\right]\left(\gamma^2(\vq_1,\vq_2,\vq_3)-\beta(\vq_1,\vq_2)\right)\\ \nonumber
&&\quad\stackrel{\text{EdS}}{=}\fr{1}{2}\left[c_{\delta,1}(a)-c_{\delta,3}(a)\right]{}^*[\d_i \delta^{(1)}(a) \frac{\d^i}{\d^2}\theta^{(2)} (a)]_{\vk},
\eea
where the expression for clustering quintessence takes the same form as for $w$CDM, and we used \eqref{tintegral}.

In the second and third lines of \eqref{fluidtaylor} we can take all fields at linear order. We have
\bea
&&\int^a \fr{da'}{a'}\; c_{\delta}(a,a')\;\frac{1}{2}\d_i\d_j \delta(x,a')\int_{a'}^a 	\fr{da''}{a''^2H(a'')}\; v^{(1)}{}^i(\vec x,a'')\int_{a'}^a 	\fr{da'''}{a'''^2H(a''')}\;v^{(1)}{}^j(\vec x,a''')\\ \nonumber
&&\quad+\int^a \fr{da'}{a'}\; c_{\delta}(a,a')\;\d_i\delta(x,a')\int_{a'}^a 	\fr{da''}{a''^2H(a'')}\; \d_j v^{(1)}{}^i(\vec x,a'') \int^a_{a''} 	\fr{da'''}{a'''^2H(a''')}\;v^{(1)}{}^j(\vec x,a''')\ \\ \nonumber
&&=\int^a \fr{da'}{a'}\; c_{\delta}(a,a')\;\frac{1}{2}\frac{D_{+}(a')}{D_{+}(a)}\d_i\d_j \delta^{(1)}\;\sfrac{\d^i}{\d^2}\theta^{(1)}\sfrac{\d^j}{\d^2}\theta^{(1)}\int_{a'}^a 	da''\frac{D'_{+}(a'')}{C(a'')D_{+}(a)}\int_{a'}^a 	da'''\frac{D'_{+}(a''')}{C(a''')D_{+}(a)}\\ \nonumber
&&\quad+\int^a \fr{da'}{a'}\; c_{\delta}(a,a')\;\frac{D_{+}(a')}{D_{+}(a)}\d_i \delta^{(1)}\; \sfrac{\d_j\d^i}{\d^2}\theta^{(1)}\; \sfrac{\d^j}{\d^2}\theta^{(1)}\int_{a'}^a 	da''\frac{D'_{+}(a'')}{C(a'')D_{+}(a)}\int^a_{a''} 	da'''\frac{D'_{+}(a''')}{C(a''')D_{+}(a)}\\ \nonumber
&&=\int^a \fr{da'}{a'}\; c_{\delta}(a,a')\;\frac{1}{2}\frac{D_{+}(a')}{D_{+}(a)}\left(\mG(a)-\frac{D_{+}(a')}{D_{+}(a)}\mG(a')\right)^2\left[\d_i\d_j \delta^{(1)}\;\sfrac{\d^i}{\d^2}\theta^{(1)}\sfrac{\d^j}{\d^2}\theta^{(1)}+\d_i \delta^{(1)}\; \sfrac{\d_j\d^i}{\d^2}\theta^{(1)}\; \sfrac{\d^j}{\d^2}\theta^{(1)}\right](a)\\ \nonumber
&& \!\!\!\!=\left(\fr{\mG(a)^2}{2}c_{\delta,1}-\mG(a)(c_{\delta_2,\mG^{\delta}_{1}}+c_{\delta_2,\mG^{\delta}_{2}})+\fr{1}{2}(c_{\delta,\mG^{\delta}_{1}}+c_{\delta,\mG^{\delta}_{2}})\right)\left[\d_i\d_j \delta^{(1)}\;\sfrac{\d^i}{\d^2}\theta^{(1)}\sfrac{\d^j}{\d^2}\theta^{(1)}+\d_i \delta^{(1)}\; \sfrac{\d_j\d^i}{\d^2}\theta^{(1)}\; \sfrac{\d^j}{\d^2}\theta^{(1)}\right](a) \\ \nonumber
&&\stackrel{w\text{CDM}}{=}c_{\delta,123}(a)\left[\d_i\d_j \delta^{(1)}\;\sfrac{\d^i}{\d^2}\theta^{(1)}\sfrac{\d^j}{\d^2}\theta^{(1)}+\d_i \delta^{(1)}\; \sfrac{\d_j\d^i}{\d^2}\theta^{(1)}\; \sfrac{\d^j}{\d^2}\theta^{(1)}\right](a)\ . \nonumber
\eea 

For completeness, the flow terms from $\delta^2$ and $s^2$ read
\bea
&&2\left(\mG(a)c_{\delta^2,1}-c_{\delta^2,\mG^{\delta}_{1}}-c_{\delta^2,\mG^{\delta}_{2}}\right)[\delta^{(1)}\d_i\delta^{(1)}\frac{\d^i}{\d^2}\theta^{(1)}]_{\vk}(a)\\ \nonumber
&&\stackrel{w\text{CDM}}{=}2 c_{\delta^2,12}[\delta^{(1)}\d_i\delta^{(1)}\frac{\d^i}{\d^2}\theta^{(1)}]_{\vk}(a)\ ,
\eea
\bea
&&2\left(\mG(a)c_{s^2,1}-c_{s^2,\mG^{\delta}_{1}}-c_{s^2,\mG^{\delta}_{2}}\right)[s_{lm}^{(1)}\d_i (s^{lm})^{(1)}\frac{\d^i}{\d^2}\theta^{(1)}]_{\vk}(a)
\\ \nonumber
&&\stackrel{w\text{CDM}}{=}2 c_{s^2,12}[s_{lm}^{(1)}\d_i (s^{lm})^{(1)}\frac{\d^i}{\d^2}\theta^{(1)}]_{\vk}(a)\ .
\eea			

\section{Full Posteriors}\label{appendixd}

\begin{figure}[h]
	\centering
	\includegraphics[width=0.99\textwidth]{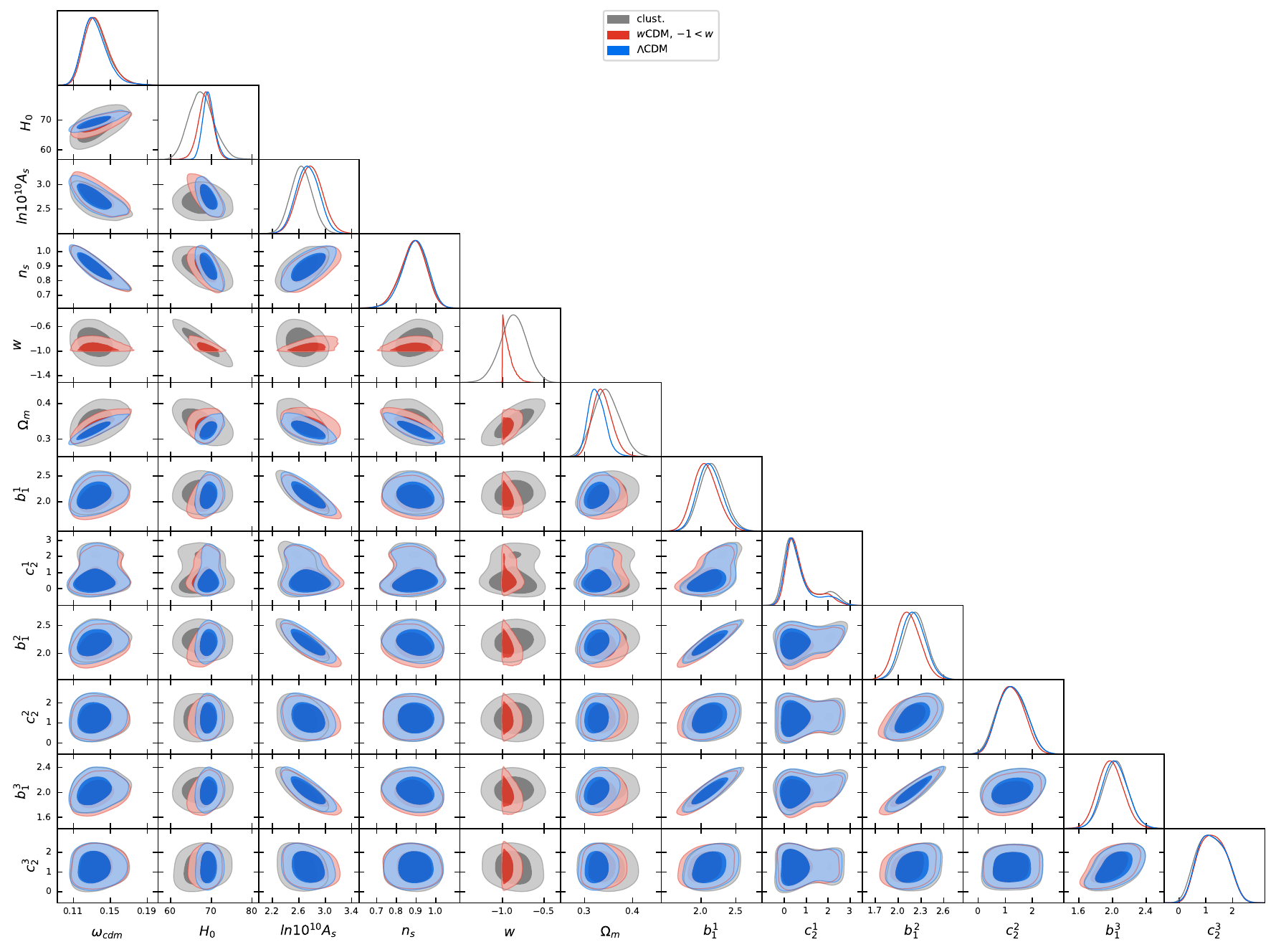}
	\caption{\small Full posteriors for the fits of clustering quintessence, $w$CDM with physical prior $w \geq 1$, and $\Lambda$CDM, to BOSS with a BBN prior.
	We show the non-analytically marginalized biases $b_1^i$ and $c_2^i$, where $i$ denotes the skycuts: $i=1$ is CMASS NGC, $i=2$ is CMASS SGC, $i=3$ is LOWZ NGC.}
	\label{fig:fulltriangle}
	\end{figure}

In fig.~\ref{fig:fulltriangle}, we show the full posteriors for all cosmological and the non-analytically marginalized bias parameters for the analysis of BOSS data alone of fig.~\ref{fig:qcdm_vs_wcdm}.

\end{appendix}

\bibliographystyle{JHEP}
\bibliography{references}

\end{document}